\documentclass[acmtog]{acmart}

\author{Arjun S. Lakshmipathy}
\email{aslakshm@andrew.cmu.edu}

\author{Nicole Feng}
\email{nfeng@andrew.cmu.edu}

\author{Yu Xi Lee}
\email{yuxilee@andrew.cmu.edu}

\author{Moshe Mahler}
\email{mmahler@andrew.cmu.edu}

\author{Nancy S. Pollard}
\email{nsp@cs.cmu.edu}

\affiliation{%
   \institution{Carnegie Mellon University}%
   \city{Pittsburgh}%
   \state{Pennsylvania}%
   \country{USA}}

\acmSubmissionID{244}

\usepackage{booktabs} 

\usepackage{graphicx, caption, subcaption, soul, xcolor}
\graphicspath{ {figures/} }

\DeclareMathOperator*{\argmin}{arg\,min}

\newcommand{\vect}[1]{\boldsymbol{#1}}

\usepackage{tabularx}

\usepackage{array}
\newcolumntype{P}[1]{>{\centering\arraybackslash}p{#1}}
\newcolumntype{M}[1]{>{\centering\arraybackslash}m{#1}}

\setcopyright{rightsretained}
\acmJournal{TOG}
\acmYear{2023}
\acmVolume{42}   
\acmNumber{4}
\acmArticle{1}
\acmMonth{8}
\acmPrice{15.00}
\acmDOI{10.1145/3592117}

\usepackage{algorithm}
\usepackage{algorithmicx}
\usepackage{algpseudocode}

\algtext*{EndProcedure}
\algtext*{EndIf}
\algtext*{EndFor}
\algtext*{EndWhile}
\newcommand{\InputConditions}[1]{\Require \parbox[t]{\dimexpr\linewidth-\algorithmicindent}{#1\strut}}
\newcommand{\OutputConditions}[1]{\Ensure \parbox[t]{\dimexpr\linewidth-\algorithmicindent}{#1\strut}}

\newcommand{\Proc}[1]{\mbox{\textsc{#1}}}

\definecolor{commentblue}{rgb}{0.643, 0.651, 0.820}
\renewcommand{\Comment}[1]{\hfill\textcolor{commentblue}{\(\triangleright\)\textit{#1}}}

\newcounter{algo}
\newenvironment{algo}[1]
{ \refstepcounter{algo}\noindent\rule{\columnwidth}{1.25pt} \\ \textbf{Algorithm~\thealgo} #1 \\ \noindent\rule{\columnwidth}{.5pt} }
{ \noindent\rule{\columnwidth}{.5pt} }

\usepackage{wrapfig}

\algnewcommand{\LeftComment}[1]{\textcolor{commentblue}{\(\triangleright\)\textit{#1}}}

\newcommand{\Csf}{\mathsf{C}}

\newcommand{\Esf}{\mathsf{E}}

\newcommand{\Lsf}{\mathsf{L}}

\newcommand{\Tsf}{\mathsf{T}}

\def\RR{\mathbb{R}} 
\def\CC{\mathbb{C}} 

\begin{document}

\title{Contact Edit: Artist Tools for Intuitive Modeling of Hand-Object Interactions}

\begin{abstract}
Posing high-contact interactions is challenging and time-consuming, with hand-object interactions being especially difficult due to the large number of degrees of freedom (DOF) of the hand and the fact that humans are experts at judging hand poses. This paper addresses this challenge by elevating contact areas to first-class primitives. We provide \textit{end-to-end art-directable} (EAD) tools to model interactions based on contact areas, directly manipulate contact areas, and compute corresponding poses automatically. To make these operations intuitive and fast, we present a novel axis-based contact model that supports real-time approximately isometry-preserving operations on triangulated surfaces, permits movement between surfaces, and is both robust and scalable to large areas. We show that use of our contact model facilitates high quality posing even for unconstrained, high-DOF custom rigs intended for traditional keyframe-based animation pipelines. We additionally evaluate our approach with comparisons to prior art, ablation studies, user studies, qualitative assessments, and extensions to full-body interaction.

\end{abstract}

%
%


%
%


\begin{teaserfigure}
\begin{center}
\includegraphics{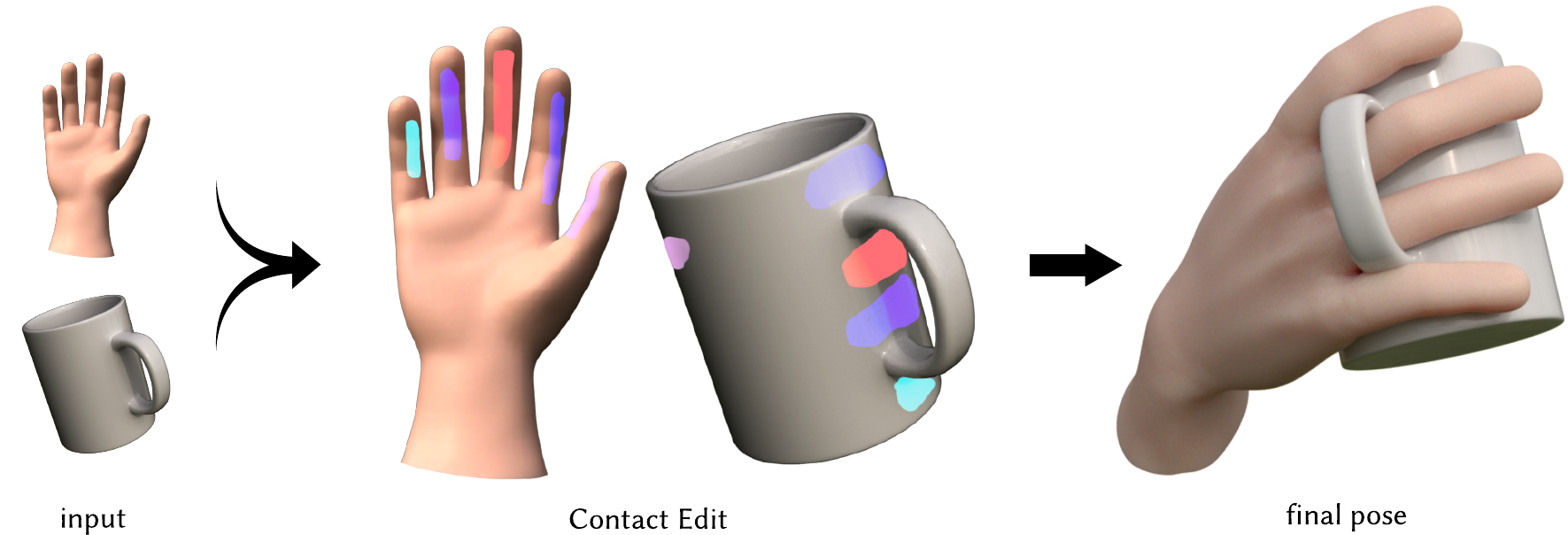}
\end{center}
\end{teaserfigure}

\maketitle

\section{Introduction}

Posing high contact interactions remains a challenging and time-consuming endeavor, even for seasoned animators. Despite the intrinsic coupling of manipulator and object motion in the physical world, production workflows today largely treat the bodies independently, requiring the animator to pose objects and all of the manipulator joints in a layered, painstaking process utilizing forward and inverse kinematics. While several recent works have endeavored to partially automate the process with exciting results (e.g., \cite{zhang2021manipnet}), these approaches are rarely adopted in traditional keyframe-based animation pipelines due to their reliance on priors, limited generalization to custom assets, and lack of a built-in feedback loop enabling detailed artistic control over results.

Consequently, manual kinematic posing remains the dominant strategy in production workflows today. This approach enables full artistic control and can work for any interaction; however, changes to the object or hand scale, geometries, kinematic structures, or even basic isometric transformations require complete re-casting of the desired poses, resulting in a laborious and costly process. We argue that the lack of intrinsic pose coupling is ultimately responsible for the majority of these difficulties. To address this challenge, we propose to model this coupling explicitly by providing artist tools to work directly with contact areas.

Contact areas, as the interaction interface between bodies in the real world, present an ideal primitive to model coupling between hands and objects. Numerous recent works have showcased the benefits of using contacts to drive grasping strategies (e.g., \cite{ye2012contactsampling,mordatch2012contact,brahmbhatt2019contactgrasp}); however, these techniques are designed more with automation in mind than with the goal of artist control. Without the ability to \emph{edit} arrangements, the value of using contacts as viable primitives rapidly diminishes.

This paper aims to address the aforementioned shortcomings, such that contacts can be integrated into production workflows directly, intuitively, and robustly. Specifically, we:

\begin{itemize}
    \item propose and demonstrate contact areas as a useful mechanism for specifying high-contact interactions,
    \item present end-to-end art-directable tools for defining contact areas, directly editing contact areas, and transferring contact areas between surfaces,
    \item define a novel contact model and foundational algorithms to manipulate and transfer contact areas using the model,
    \item modify an existing optimization framework to pose deformable, unconstrained, high DOF manipulators,
    \item will release [on publication] a Maya plugin for use and to facilitate comparisons to our approach.
\end{itemize}

The techniques proposed in this paper are fully deterministic, not reliant on external priors, computationally inexpensive, robust to high complexity interactions, provide transparent behavior, enable intuitive interfaces, and are designed to facilitate artistic interaction with the animator. We validate these claims by putting our plugin in the hands of real animators to show that high quality results can be generated even with minimal training. Finally, although this paper primarily targets hand-object interaction, we show that the approach generalizes well to other scenes featuring large areas of contact, such as a full-body character sitting down into a chair.

\section{Production Process Overview}

Modern production workflows for traditional 3D keyframe-based CG media involve a large number of stakeholders but generally follow a standardized pipeline depicted below. More comprehensive overviews are available online\footnote{https://dreamfarmstudios.com/blog/3d-animation-pipeline/}:
\begin{figure}[h!]
\centering
\includegraphics[width=\linewidth]{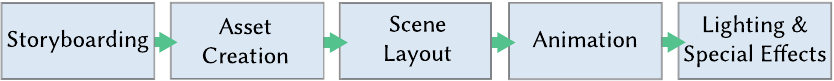}
\end{figure}

Each stage is carried out by one or multiple subteams and overseen by an art director, while the entire production is overseen by an executive director. While the pipeline is effective in promoting specialization and is scalable to large organizations, the separation of work does create several important caveats in operations.

Character assets are modeled, rigged, and skinned at the asset creation stage, and are subsequently delivered as final products to the animation stage. Animators do not typically modify these assets, but nonetheless are expected to generate high quality poses and must be able to iterate and explore multiple poses while including input from the art and executive director. The rigging team will create basic and unconstrained control handles to address the need for grasp oriented interaction. These handles are often generic and may not consider where and how the hand will make contact with the object, creating substantial complexity for the animator. Even if a studio provides custom tools to add control handles in an ad-hoc manner, the animator will still need to manipulate the skeletal rig, approximating how the surface meshes of both the object and the hand will interact. The end result is a process which is especially challenging, highly manual, and time-consuming for poses involving multi-asset interaction such as hands and objects. These tasks are particularly daunting for early-career animators, who often end up ``fighting the rig" due to their lack of experience.

One potential solution is to enable animators to define interactive poses on meshes directly and instead compute the optimal corresponding rig pose automatically; however, it is \textit{equally} challenging to create robust tools which ``just work" on a wide variety of geometries due to fundamental differences in mesh desiderata between the geometry processing and animation communities. For example, the tight coupling between rigging and skinning biases asset modeling towards the construction of coarse meshes of quadrilateral primitives, while many geometry processing algorithms assume, if not outright require, triangle primitives. Although quads can be trivially triangulated, there is no oversight to ensure that these assets have additional desirable properties (e.g. uniform sampling, no triangle inequality violations, etc.). Additionally, in an effort to reduce asset creation time and overall production costs, a growing number of assets today are obtained from alternate sources such as 3D scans or CAD models designed for printing, which have far more degenerate properties than in-house artist models.

\textit{Art-directable (AD)} tools are expected to operate under such workflows and provide an easy-to-understand interface, produce predictable behavior, enable refinement, be reasonably fast and robust to a wide range of assets, and must be flexible enough to accommodate a wide variety of user and project needs in order to provide value. \textit{End-to-end art-directable (EAD)} tools, which enable animators to create results from scratch entirely under the tool framework, are even more challenging to design; instead, the large majority of tools today are ``last-mile" only, requiring the animator to perform a significant amount of upfront work manually in order to utilize them. This paper presents EAD tools for hand-object pose generation through the direct manipulation of contact areas, which are fully compatible with existing posing tools, founded on principled mathematics, easy to understand, and more broadly generalizeable to multi-object interactions.

\section{Related Works}

Contact-driven approaches have been employed on numerous occasions in a variety of contexts. We consider three broad categories of related prior work: pose refinement; grasp and manipulation synthesis; and inverse kinematics. We also survey the texture and material transfer literature, which is relevant to our particular method.

\subsection{Pose Refinement}

One of the most popular uses of contact information is to refine coarse poses, often with the goal of removing unwanted artifacts and improving physical plausibility. For example, contact points have been used to drive physics-based motion synthesis \cite{liu2008synthesis}, refine tracked data into physically plausible solutions \cite{ye2012contactsampling}, generate hand motion for full body motion capture through stochastic optimization \cite{zhao2013physicscontrol}, and improve the quality of hand-hand interactions through compiled databases \cite{lee2017multifinger}. They have been used in some interactive applications, in one prior work enabling animators to specify contact points to facilitate fusing of independently tracked hand and object motion \cite{hamer2011ddhoi}. All of these approaches use single point contact models.

Area and volumetric representations of contact provide greater solution consistency over larger regions. For example, deformable models have been utilized to correct reconstructed geometries in an effort to resolve self-collisions \cite{smith2020elasticity}. Regional contact maps have also been inferred \cite{grady2021contactopt, jiang2021graspTTA} by training on publicly available datasets \cite{brahmbhatt2020contactpose, brahmbhatt2019contactdb} or in house using motion capture \cite{taheri2020grab} to optimize coarse pose estimates. Like these works, we exploit full contact regions, but address the scenario where ground truth data is not available and may not exist. Instead, we strive to provide the animator with tools to create and manipulate contact regions to create an envisioned result from scratch.

\subsection{Grasp Synthesis}

Contacts have played a vital role in the generation of grasps and manipulations. Computing kinematically or dynamically feasible whole hand grasps, especially for dexterous manipulators, is a notoriously challenging problem with a long studied history. Historical works typically considered grasps analytically by examining point contact models with Coulomb friction \cite{murray1994manipulation, siciliano2007handbook}. Other techniques include performing global searches for kinematically feasible grasps meeting certain constraints \cite{pollard1997parallelgrasps}, leveraging passive compliance to exploit automatic contact dynamics from physics simulation \cite{pollard2005physicalgrasping}, and targeting optimal independent regions within which point contacts can be placed \cite{roa2009icr3d}. Contacts have also been used as error terms in the training of generative models \cite{christen2021dgrasp, wu2021SAGA}.  A number of works target manipulation planning and optimization by switching between contact modes \cite{cheng2021contactmode2d} or by optimizing contact placement and forces for physically plausible results \cite{mordatch2012contact,hazard2020automated}. The $\epsilon$ metric proposed by Ferrari and Canny for evaluating grasp wrench spaces \cite{ferrari1992metric} -- the most widely used grasp quality evaluation metric -- is fundamentally rooted in its analysis at individual contact points.

However, characterization of grasps as single points highly simplifies the complexities of real interactions and cannot account for geometric consistency outside the designated point. This is especially problematic for manipulators with deformable skins, which may utilize large contact regions to grasp objects. Solutions capable of generating high quality results through the integration of differentiable contact dynamics have been presented \cite{turpin2022graspd}; however, it is challenging to define a custom grasp through this framework or compute solutions in timescales acceptable for interactive iteration. Instead, the precedents of optimizing poses by matching contacts \cite{brahmbhatt2019contactgrasp, lakshmipathy2022contacttransfer} are most relevant to our work; however, these methods focus largely on grasp generation and assume that contact areas defining the grasps are available from external sources. This paper provides the critical missing pieces required to create and edit contact areas from the ground up, thus eliminating reliance on external data completely and enabling the creation of EAD tools. We also introduce a simple, but necessary modification to the optimization formulation presented in previous work to enable our tools to compute solutions for unconstrained high DOF rigs.

\subsection{Inverse Kinematics}

Posing characters to meet contact targets has long been addressed through inverse kinematics (IK).  Zhao and Badler were the first to describe inverse kinematics as a general optimization problem \cite{zhao1994inverse}, pointing out the flexibility of formulating optimal pose with an objective function rather than following a specific iterative IK process. This theme has been carried forward in follow-up works which allowed artists to work with many IK handles stably and robustly to pose characters \cite{yamane2003natural}. A relatively recent survey of IK approaches can be found in the work of Harish et. al. \cite{harish2016parallel}. Our optimization can be viewed as a solution to the IK problem when the IK targets are contact areas. We present a modification of the optimization approach in previous works \cite{lakshmipathy2022contacttransfer} that works reliably and robustly to solve this problem.

\subsection{Texture and Material Transfer}

Our method of contact parameterization and transfer based on logarithmic maps (i.e. geodesic distance and direction) is similar to methods previously used in texture transfer and mesh editing. Like our method, previous works use the log map to parameterize surface patches with respect to a ``spine" and allow user control over the transfer process by specifying a starting point and direction for the spine on the target mesh \cite{Biermann:2002:cutpaste}. However, the existing technique \cite{Biermann:2002:cutpaste} requires both the source and the target mesh to be parameterized over the plane, which renders the method complicated and costly. In contrast, our method performs no intermediate parameterizations: contact patches are instead parameterized directly via surface-computed log maps and patches are transferred by tracing geodesics on the target surface. The log map parameterization has also been used to transfer texture ``decals" \cite{schmidt2006dem}; however, the computed log maps are approximations and exhibit accuracy degeneration that depends on the sampling of the underlying surface. This technique \cite{schmidt2006dem} also experiences texture ``foldover'' when the underlying sampling is too sparse, where discretization artifacts cause the texture fold over itself. We instead compute log maps which are always exact in the polyhedral sense using the MMP algorithm \cite{MMP:1987} while still maintaining interactive rates. Additional control is provided by allowing users to parameterize contact patches with respect to an axis of choice, enabling artifacts to be avoided even for difficult transfers.  In summary, our approach contributes interactive texture transfer and editing, with log maps that are always exact in the polyhedral sense, and that avoids the need for an intermediate mesh parameterization as well as the high patch distortions and foldover that can come from prior approaches. We provide the artist with the ability to define and change their own axis , but also provide a default axis to start with.

Some recent research also demonstrates flexible use of local log map parameterizations for contact, but defines the contacts differently, including based on a boundary or based on a single representative point~\cite{lakshmipathy2021contacttracing, lakshmipathy2022contacttransfer}. Neither existing representation was effective for our purposes, and we compare to them directly in Section~\ref{sec:contactModelComparisons}.

Our method of designating transfer via curves also bears some similarity to existing work on cloth layering for arbitrary bodies via an ordered set of ``freedom marks" \cite{igarashi2002clothingtransfer} as well as determining equivalent functional maps \cite{gehre2018interactivecurve}. Our method can be viewed as extending the literature on designating correspondences via input curves, in this case moreso as a basis for IK.

\section{Methods}
\label{sec:sim}

\begin{figure}[t]
    \centering
    \includegraphics[width=1.0\linewidth]{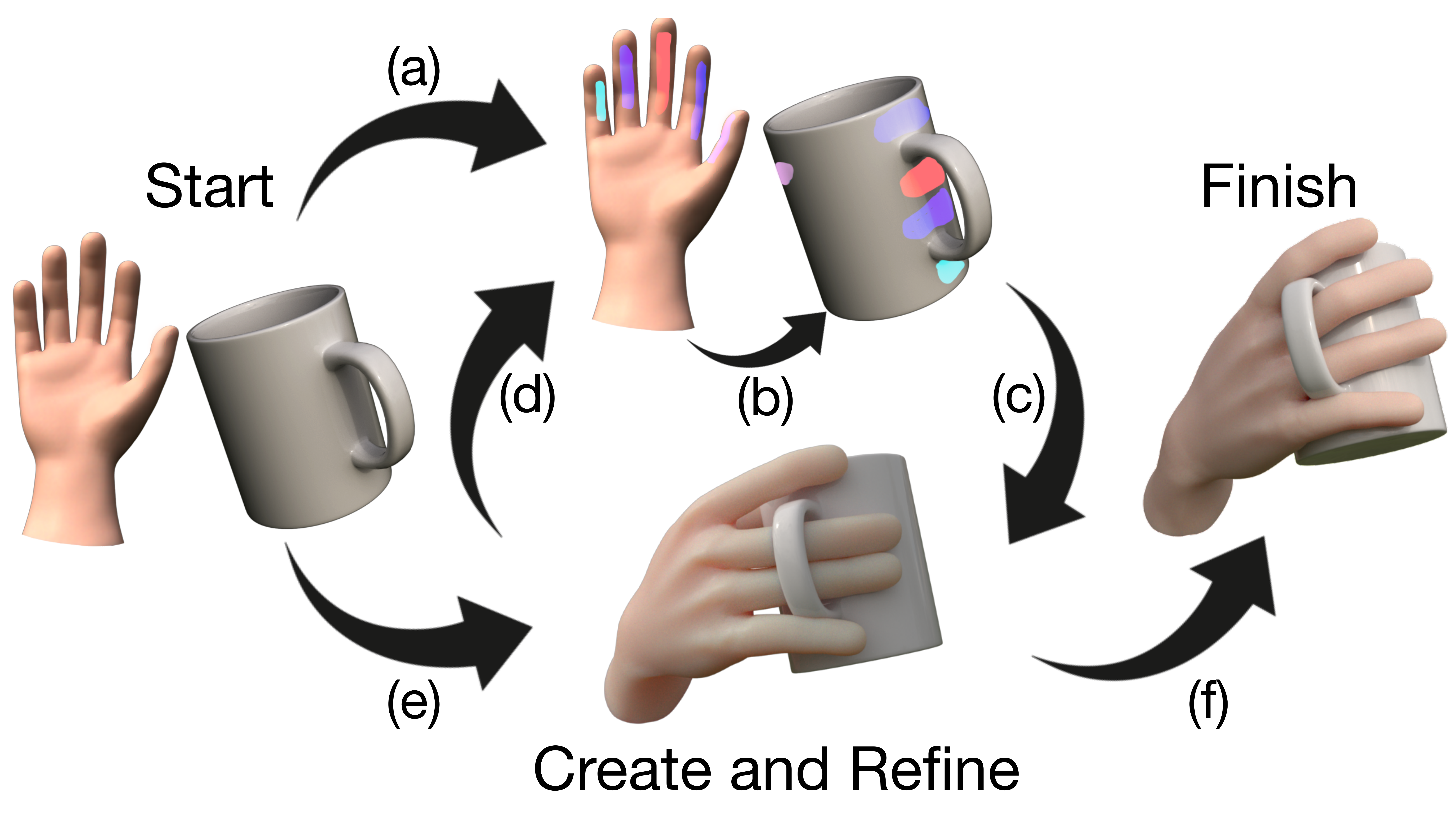}
    \caption{High level overview of our tool workflow. Starting from a given hand and object, the artist typically starts by (a) creating contact areas on the hand and manipulating them into their final positions before (b) transferring them to the object to generate a corresponding set. Artists can then (c) compute the manipulator pose from the contact areas and (d) use the result as feedback to edit the contact areas on the hand and object in an iterative feedback loop. Artists can alternatively (e) first create a (often coarse) pose using traditional posing tools and designate contact areas afterword. (f) Refinement continues until the final result is acceptable.}
    \label{fig:workflow}
\end{figure}

Figure \ref{fig:workflow} illustrates the workflow we aim to enable, which requires the ability to create and manipulate contact areas, transfer areas between surfaces, and compute the appropriate pose. To support this workflow, we present our contact model, define the operations supported by our model, and detail the optimization process used to compute character pose from contact area correspondences.

\begin{figure*}[t]
    \centering
    \includegraphics[width=1.0\linewidth]{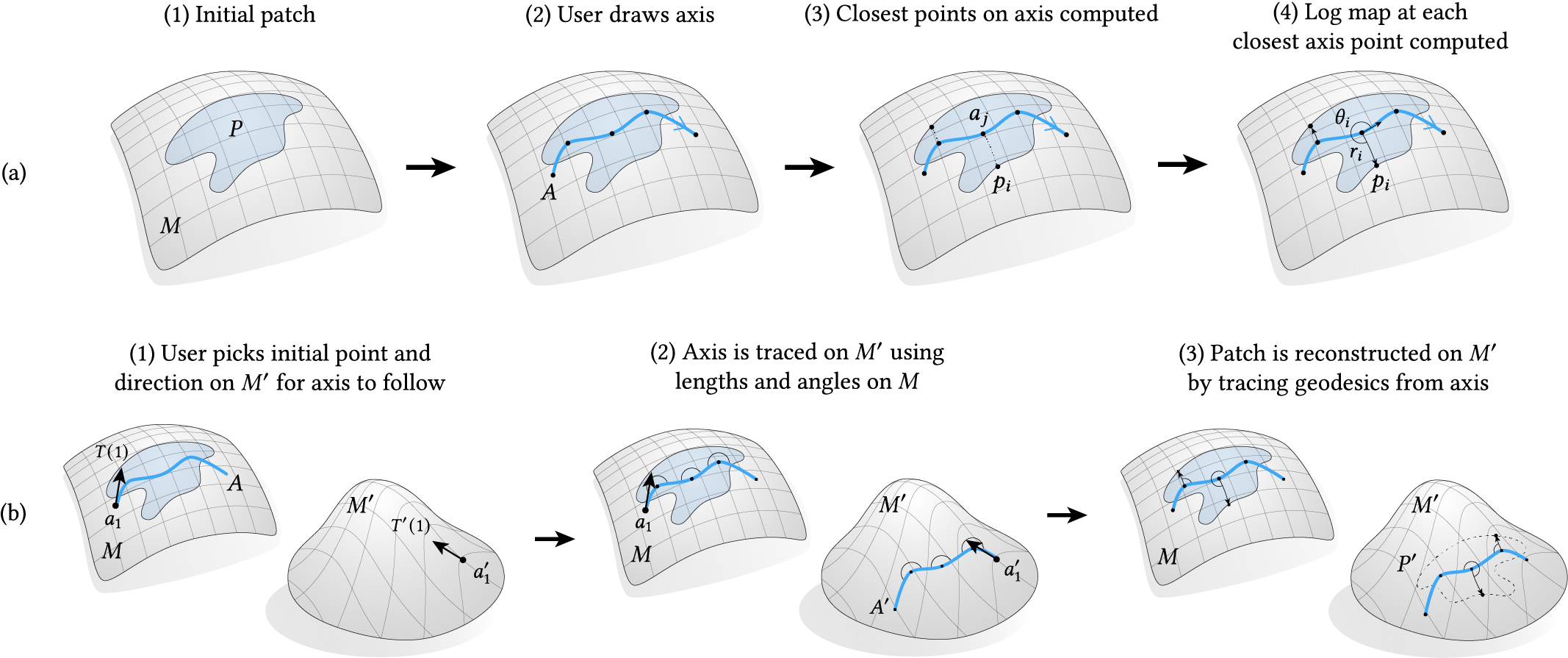}
    \caption{An overview of contact patch parameterization (a) and patch transfer (b). See Section~\ref{sec:contact-model} for technical detail.}
    \label{fig:contact-transfer-diagram}
\end{figure*}

\subsection{Contact Parameterization \& Transfer}
\label{sec:contact-model}

\subsubsection{Smooth formulation}

We define a contact patch $\mathcal{P}\subset\Omega$ as a closed region on surface $\Omega$. To enable direct editing of contact patches, we must develop a representation that enables patch \emph{transfer} and patch \emph{manipulation}. \emph{Transfer} is the task of computing an as-isometric-as-possible embedding of a patch $\mathcal{P}$ on a source surface $\Omega$ into another target surface $\Omega'$. Patch \emph{manipulation} includes any isometric transformations to $\mathcal{P}$ (translation, rotation, etc.) Importantly, our representation must enable interactions with patches to be intuitive, predictable, and fast.

Our first insight is that embedding an open curve is significantly easier than embedding a region or closed loop. Hence we parameterize $\mathcal{P}$ with respect to an open curve $\mathcal{A}$ which we call an ``axis''. We define axis $\mathcal{A}$ as a piece-wise geodesic curve on $\Omega$ containing:

\begin{enumerate}
    \label{enum:axis-description}
    \item a finite set of points $\{a_1,\dots,a_m\}\in\Omega$, with a \emph{shortest geodesic} $g_i$ connecting each pair of adjacent points $(a_i, a_{i+1})$ for $i=1,\dots,m-1$;
    \item \emph{turning angles} $\{\phi_i\}_{i=2}^{m-1}$, where each $\phi_i$ is the angle of rotation from the ending direction of $g_{i-1}$ to the initial direction of $g_i$, expressed in the tangent space of $a_i$.
\end{enumerate}

\noindent For every point $p\in\mathcal{P}$, we store:

\begin{enumerate}
    \item the closest point $q$ on $\mathcal{A}$
    \item the \emph{logarithmic map} $\log_q(p)$ encapsulating the distance $r_p$ from $q$ to $p$ and the initial direction $\theta_p$ of a geodesic from $q$ to $p$ expressed in the tangent space of $q$. The angle $\theta_p$ is taken relative to the tangent direction $T(q)$ of $\mathcal{A}$ at $q$. (If $q = a_i$ for some $i$, then $T(q)$ is taken as either the initial direction of $g_i$, or the ending direction of $g_{i-1}$.) See inset for diagram.
\end{enumerate}

\setlength{\columnsep}{1em}
\setlength{\intextsep}{-1em}
\smallskip
\begin{wrapfigure}{r}{83pt}
\centering
\includegraphics{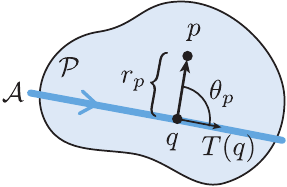}
\label{logmap_diagram}
\end{wrapfigure}

With this parameterization, $\mathcal{P}$ can be transferred from $\Omega$ to $\Omega'$ by first transferring the axis $\mathcal{A}$. In what follows, let primed quantities denote the corresponding transferred quantity on $\Omega'$. To transfer $\mathcal{A}$, the position of its first point and initial tangent direction must be specified on $\Omega'$; then the position of $\mathcal{A'}$ on $\Omega'$ is completely determined by re-tracing its constituent geodesic segments on $\Omega'$. The transferred patch $\mathcal{P}'$ can then reconstructed on $\Omega'$ by simply applying the \emph{exponential map}, the inverse of the log map: For each $p\in P$, the corresponding $p'\in \mathcal{P'}$ is obtained by tracing a geodesic starting from $q'$ at an angle $\theta$, taken relative to the tangent of $\mathcal{A'}$ at $q'$, for a distance $r$.

The reason for parameterizing points $p\in\mathcal{P}$ relative to closest points in $\mathcal{A}$ is so that $\mathcal{A}$ acts as an intuitive control handle roughly corresponding to the ``skeleton'' of $\mathcal{P}$. Using closest points also distributes these ``basis points'' along $\mathcal{A}$, which prevents the foldover that might occur in the transferred patch if we only parameterized $P$ with respect to a single point (see Figure~\ref{fig:logmapcomparison}.)

In general, the embeddings that result from this parameterization will not be truly isometry-preserving; for example, a geodesic ball in a region with positive curvature will have a smaller area than a geodesic ball of the same radius in a region of zero curvature. The embedding will be isometry-preserving if and only if the source region on $\Omega$ and target region on $\Omega'$ are themselves isometric, which will almost never occur. However, the strength of this method is that it can yield a reasonable embedding that approximately preserves the shape of the patch, whether or not an isometric embedding exists. Such robustness is especially important in artist-driven workflows, where physical and geometric plausibility is usually not strictly followed. Furthermore, the majority of surface regions used as contact regions in practice have low curvature, which inherently minimizes distortion.

The parameterization is also robust to the choice of axis. The exponential map is surjective by the Hopf-Rinow theorem, meaning any point $p$ on a connected surface $\Omega$ can be reached from any other point $q\in \Omega$ by following a geodesic from $q$ for the appropriate distance with the appropriate initial direction. Thus a parameterization can be computed for any choice of axis (although some axes will be more intuitive or yield better parameterizations than others -- see Section~\ref{sec:DrawbacksAndLimitations}).

\subsubsection{Discrete implementation}

Parameters and the transfer process for the discrete formulation are illustrated in Figure~\ref{fig:contact-transfer-diagram}. We discretize the surfaces $\Omega$, $\Omega'$ as manifold triangle meshes $M$ and $M'$ respectively, both of which may or may not have boundary. Analogous to our setup in the smooth setting, the discrete axis $A$ is specified by:

\begin{enumerate}
    \label{enum:discrete-axis-description}
    \item a finite set of points $\{a_1,\dots,a_m\}$ on $M$, with a discrete geodesic $g_i$ connecting each pair of adjacent points $(a_i, a_{i+1})$ for $i=1,\dots,m-1$;
    \item turning angles $\{\phi_i\}_{i=2}^{m-1}$, where each $\phi_i$ is the angle of rotation from the ending direction of $g_{i-1}$ to the initial direction of $g_i$, expressed in the tangent space of $a_i$.
\end{enumerate}

\noindent The patch is represented as a set of barycentric points $P = \{p_1,\dots,p_n\}$ on $M$. For every point $p_i\in P$, we will compute:

\begin{enumerate}
    \item the closest point $a_j\in\mathcal{A}$ to $p_i$;
    \item the log map $\log_{a_j}(p_i)$, which represents the geodesic distance $r_i$ from $a_j$ to $p_i$, and the initial direction $\theta_i$ of a geodesic from $a_j$ to $p_i$, expressed in the tangent space of $a_j$. The angle $\theta$ is taken relative to the tangent direction $T(a_j)$ of $\mathcal{A}$ at $a_j$. $T(a_j)$, also denoted $T(j)$, is defined as the initial direction of the geodesic connecting $a_j$ to $a_{j+1}$, expressed in the tangent space of $a_j$; for the axis endpoint $a_m$, we use the direction from $a_m$ to $a_{m-1}$ instead.
\end{enumerate}

In our plugin, the user need only specify the sequence of points $\mathcal{A}$, while $\{g_i\}_{i=1}^{m-1}$ and $\{\phi_i\}_{i=2}^{m-1}$ are computed from $A$ automatically. More optimally, a typical use has the artist select only the start and end of the axis with all edge crossings along the geodesic connecting those two points automatically added to the axis representation, minimizing requirements for user input while simultaneously providing a dense representation. Tangent vectors are represented as complex numbers in a local coordinate system. Defining a basis for tangent spaces at points on faces and edges is relatively straightforward, since each face and pair of adjacent faces is on its own intrinsically flat; we define tangent spaces at vertices using the definition of previous works \cite{sharp2019vhm}. Importantly, the definition of tangent bases at any discrete component enables our method to be \textit{sampling-agnostic}, which we demonstrate enables smooth manipulation equally well on both coarse and fine meshes. To evaluate the exponential map, we trace discrete geodesics using the scheme described by \cite{Polthier:2006:geodesics}.

To parameterize patch $P$, we first compute the closest point in $A$ to each point $p_i\in P$. Determining the closest point with absolute accuracy is not essential since the closest point only acts as the point at which the log map will be computed and does not affect the accuracy of the log map itself. In our experience, we found that it sufficed to only consider the points $\{a_1,\dots,a_m\}$ as closest points. Hence, we use the fast but approximate Vector Heat Method \cite{sharp2019vhm} to do closest-point interpolation, which determines the closest point $a_j\in A$ to each point $p_i\in P$.

For each such pair $(p_i, a_j)\in P\times A$, we then compute $\log_{a_j}(p_i)$. Since the log map accuracy directly affects the accuracy of the patch reconstruction, we use an exact polyhedral geodesic algorithm \cite{MMP:1987}, referred to here as MMP, to explicitly compute the exact geodesic $h_i$ from $a_j$ to $p_i$. We compute $\log_{a_j}(p_i)$ by computing the initial direction $\theta_{i}$ of $h_i$ in the tangent space of $a_j$. In general, $\theta_{i}$ may be expressed with respect to an arbitrary polar axis; we express $\theta_{i}$ with respect to the outgoing direction $T(j)$ of $g_j$ at $a_j$, i.e. we store $\tilde{\theta}_{i} = \theta_{i}/T(j)$. We store $\log_{a_j}(p_i)$ as a single complex number $z_i:=r_ie^{i\tilde{\theta_i}}$, where $r_i$ is the length of $h_i$. Figure \ref{fig:contact-transfer-diagram}(a) illustrates the process of patch parameterization.

We must also encode the axis $A$ for reconstruction on $M'$. For each pair $(a_i, a_{i+1})$ of subsequent points in $A$, we store the length of the geodesic segment $g_i$ connecting them. For each point $a_i$ for $i\in\{2,\dots,m-1\}$, we also store the turning angle $\phi_i$ at $a_i$, i.e. the angle between the incoming direction of $g_{i-1}$ at $a_i$, and the outgoing direction of $g_i$ at $a_i$, expressed in the tangent space of $a_i$. These lengths and angles determine the position of $A$, up to an overall translation and rotation which must be stored explicitly.

We now describe the process of patch transfer. In what follows, primed quantities denote quantities on $M'$ that have been transferred from $M$. To transfer $A$ to the target surface $M'$, the user selects a point $a'_1\in M'$. The user also specifies the initial direction $T'(1)$ of the first geodesic segment $g'_1$ of the transferred axis $A'$. The axis $A$ is transferred simply by tracing out its constituent geodesics on $M'$, starting from $a'_1$ and using $T'(1)$ as an initial direction; the rest of the axis is traced using the lengths and turning angles between geodesic segments computed from $A$ on $M$. After the axis has been transferred, the transferred patch $P'$ is reconstructed on $M'$ by tracing geodesics from the patch parameterization, subject to one caveat. In the context of transfer, we actually want $P'$ to be the \textit{mirror image} of $P$ since $P\subset M$ and $P'\subset M'$ are intended to be in contact with each other with opposing normals. Hence, the angles used for geodesic tracing during axis and patch transfer will be negated compared to the angles computed for their respective parameterizations on $M$. Figure \ref{fig:contact-transfer-diagram}(b) illustrates the process of patch transfer.

\begin{figure}[H]
    \centering
    \includegraphics[width=0.8\linewidth]{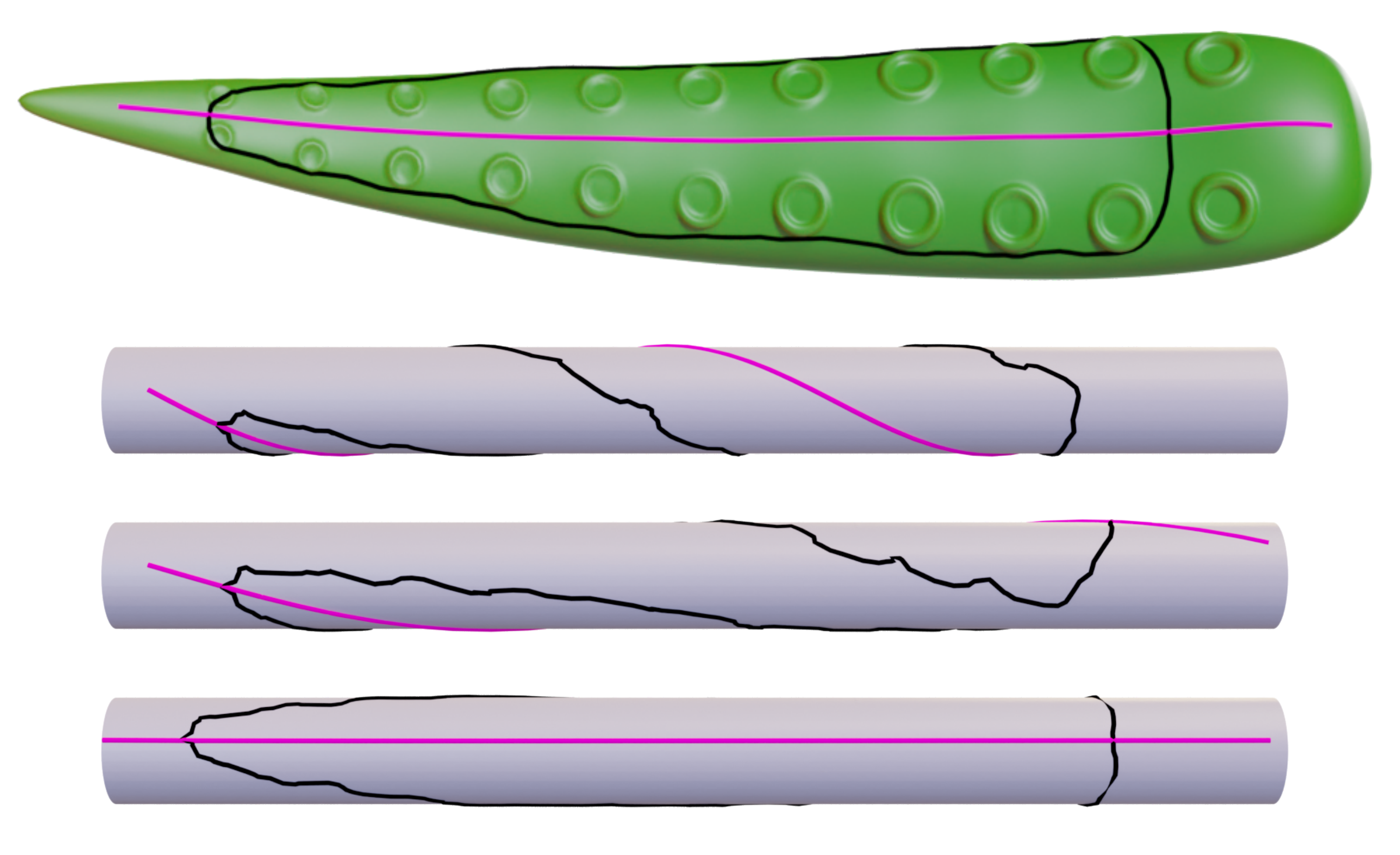}
    \caption{By drawing an appropriate axis (magenta), the user can easily specify that the patch on the tentacle should be transferred to wrap around a cylindrical object.  Changing the start angle of the axis changes the wrap.}
    \label{fig:transfer-longitudinal}
\end{figure}

As discussed in Section~\ref{sec:contact-model}, users are free to designate a variety of axes $A$. This flexibility jointly makes interaction with contacts very robust to user input and provides fine-grained control over where and how the patch should be embedded into the target surface. For example, if the user wanted the tentacle in Figure \ref{fig:transfer-longitudinal} to wrap around a cylindrical object, one intuitive option would be to draw the axis to coincide with the longitudinal axis of the tentacle patch. Choosing different initial directions for that axis creates different wraps.

Our parameterization allows us to reconstruct complex contact patch shapes exclusively through computation of geodesics and geodesic path tracing, which depends only on the mesh being manifold. Even after transfer, users can easily manipulate the patch on $M'$ simply by manipulating the axis $A'$; the patch is simply re-constructed upon any changes to the axis. Parameterization is the most expensive step of our method due to its reliance on an exact polyhedral method; however, it is a one time process and tends to not take very long in practice. Please see the supplementary video for approximate wall clock timescale on an example. Patch transfer and manipulation only requires the inexpensive operation of geodesic tracing and can be performed in real-time. We expand upon the supported contact patch editing operations in proceeding subsections and provide pseudcode for all algorithms in the appendix.

\subsection{Contact Painting}

Contacts can be created either using off-the-shelf brush tools already provided by Maya or by our custom paint brush. We leave the choice up to users but briefly discuss use cases for both options.

On fine meshes or regions with dense sampling, users may find it beneficial to select existing discrete vertices. Vertex selection is flexible, inexpensive, has no parameters beyond the brush size, and is a method familiar to most artists. The drawback, however, is that this method does not enable selection of arbitrary points such as those within faces or in between edges, which leads to problems creating dense contacts on coarse meshes.

In response, we created our own brush tool which enables arbitrary point selection by discretizing a NURBS curve drawn in screen space and projecting the points into mesh space via standard ray-mesh intersection. Our custom brush offers a solution to the coarse mesh problem; however, it requires users to designate possibly unintuitive parameters such as the discretization step size and does not provide variable brush stroke sizes.

For these reasons, we allow heterogeneous creation of contacts using both brushes, which combined permit users to work directly with the original desired triangulated mesh, even if the mesh is irregularly sampled or has poor triangulation in general.

\subsection{Contact Editing}

\begin{figure}
\centering
\includegraphics[width=1.0\linewidth]{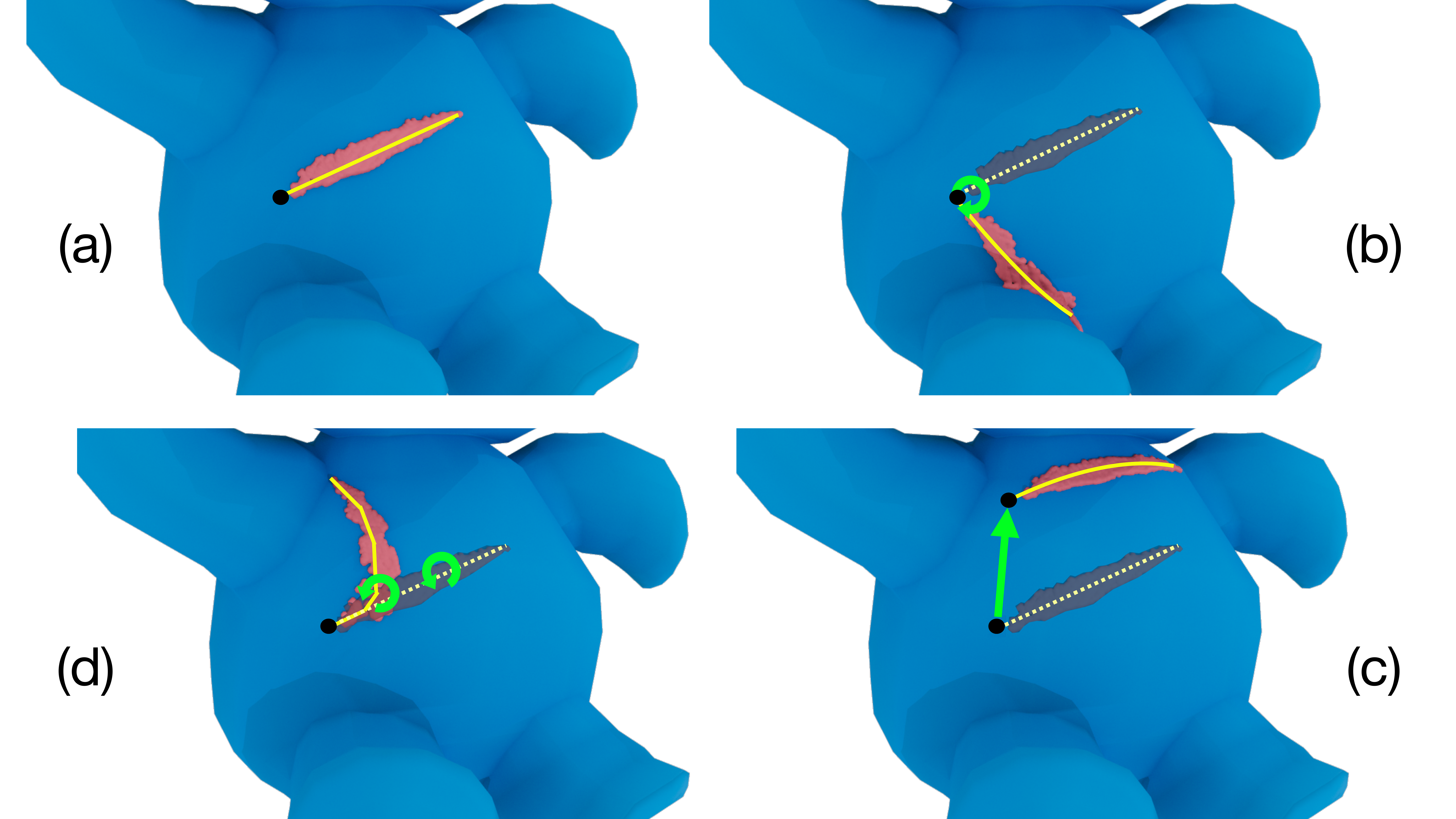}
\caption{Illustrations of our editing operations applied to a sample contact. (a) Starting from a contact painted on a coarse mesh using our custom brush, users can (b) rotate, (c) translate, or (d) locally deform the patch directly on the source domain. Black dots indicate axis start points.}
\label{fig:editingfigure}
\end{figure}

Our model supports three editing operations which aim to be as isometric as possible: translation, rotation, and transfer. We additionally support local deformation (bending) of the patch, as well as hierarchical composition by linking multiple axes together. Since the location of the patch is determined by its parameterization with respect to its axis, to edit the patch we need only edit the axis. Defining all patch editing operations as operations on the axis simplifies the editing process conceptually and computationally. Users can directly manipulate large and arbitrarily shaped contact regions using only the axis as a simple, low-dimensional control handle. All patch editing operations, which can be viewed in the supplementary video, execute at interactive rates. Pseudocode for each operation is available in the appendix.

\subsubsection{Translation}

We define patch translation as rigidly moving a patch along a specified path with as little rotation as possible with respect to the path. Mathematically, this definition corresponds to \emph{parallel transport} of the patch along the path of translation. In practice, we found that simply transporting the patch along the shortest geodesic connecting the start point to the end point of the path of translation produced more predictable behavior than parallel-transporting the patch along the user-drawn path of translation. The process of patch translation is implemented as follows:

\begin{enumerate}
\item the user drags the patch across the surface;
\item the shortest geodesic between the start and end point of the dragging path is computed using MMP;
\item the initial direction $\vec{d'}$ at the first point $a'_1$ of the patch axis is transported along this shortest geodesic, using \emph{discrete parallel transport}~\cite{sharp2019vhm};
\item the axis is re-traced at the end point, and finally the patch is reconstructed using its axis-based parameterization.
\end{enumerate}

In the third step, discrete parallel transport of a vector with respect to a path involves applying a change-of-basis to the vector to keep it parallel as it passes through different local coordinate systems, which amounts to a sequence of simple complex arithmetic operations. See Section 5.2 in \cite{sharp2019vhm} for details.

\subsubsection{Rotation}

Rotation is the task of rotating the whole patch rigidly around a designated pivot point. Rotating the patch around $a'_1$ amounts to simply choosing another direction for initial direction $d'$ of the transferred patch axis $A'$. Although it is possible to similarly define rotation around any other point in $A'$ by constraining the axis tangent direction at that point, the pivot point is always chosen to be $a'_1$ for simplicity.

\subsubsection{Local Deformation}

While the editing operations defined so far are intended to be as-rigid-as-possible transformations acting on the whole patch, we define local deformation as transforming only part of the patch rigidly, which corresponds to rotating part of the axis $\{a_i,a_{i+1},\dots,a_m\}\subset A$ around the point $a_i\in A$ while keeping the rest of $A$ fixed. Like all other patch editing operations, local deformations do not change the connectivity of the axis. Unlike global editing operations, however, this local rotation requires that we must update the turning angle $\phi_i$ at $a_i$.

\subsubsection{Hierarchical Composition}

We allow the user to edit a system of multiple patches in a hierarchical manner, by ``parenting'' (possibly multiple) child patches to a parent patch. For each child patch, we use MMP to connect its axis to the axis of its parent with a geodesic. We choose to connect the first points in the axis for simplicity, but note that connections can alternatively be made from the end point or any intermediate point if desired. The connected axes are equivalent to a single axis, with the only difference being that axis points may now possess multiple successors in the case of branching. We store the turning angles and segment lengths corresponding to each newly-added axis point. Therefore, whenever the axis of the parent is updated, we can re-trace each child axis by applying the usual axis-tracing procedure to the single connected axis. Using the connected axes ensures that the position of each child patch relative to its parent patch is approximately maintained throughout any operations applied to the parent patch.

\subsection{Kinematic Hand Posing}

After corresponding contacts on the object and hand have been created and paired, we adopt a modified formulation of the optimization in \cite{lakshmipathy2022contacttransfer} to compute the hand pose that best matches the contacts, while simultaneously accounting for surface geometries and user edits. Specifically, we start with the existing formulation:

\begin{equation}
    \begin{array}{rrclcl}
        \displaystyle \vect{\theta}^* = \argmin_{\vect{\theta}} & \multicolumn{3}{l}{\sum_{i=0}^{N}\ \  (\lambda_d \Gamma_{D,i} + \lambda_n \Gamma_{N,i}) + \sum_{j=0}^{J}\ \lambda_p \Gamma_{P,j}}\\
        \mathrm{s.t.} & \vect{\theta_L}  \leq \vect{\theta} \leq \vect{\theta_U}\\
    \end{array}
    \label{eq:optOrig}
\end{equation}

\noindent where $\vect{\theta}$ is the degree of freedom vector, N is the total number of corresponding contact vertices, $J = |\vect{\theta}|$, $\vect{\theta_L}$ and $\vect{\theta_U}$ define the lower and upper bounds of $\vect{\theta}$ respectively, $\Gamma_{D,i}$ penalizes for distance from contacts, $\Gamma_{N,i}$ attempts to align opposing surface normals, and $\Gamma_{P,j}$ encourages solutions near a rest pose or artist-defined prior.

Unfortunately, applying Eq. \ref{eq:optOrig} as-is \textit{does not work} for unconstrained, high-DOF systems characteristic of standard animation rigs. Unlike robot hands whose DOFs are constrained by real-world considerations, animation rigs are intended to provide as much artistic flexibility as possible. A common approach is to model all child joints as full 3-DOF ball joints and the root as a full 6-DOF joint. Note that production rigs are typically far more complicated than this formulation, often sporting more complex IK and NURBS handles. Even under this simple formulation, however, we found that adopting Eq. \ref{eq:optOrig} as-is produced mediocre results with highly undesirable characteristics such as contorted fingers and significant self-intersections. See Figure~\ref{fig:ablationstudy}c for one such example.

To address this problem, we make two modifications. First, we drop the root DOF contributions to $\Gamma_{P}$. Second, we select a much higher $\lambda_p$ than described in the literature. By introducing the aforementioned modifications, we are able to reliably position the manipulator near the contacts while restraining alterations to the rest pose. The revised formulation is thus:

\begin{equation}
    \begin{array}{rrclcl}
        \displaystyle \vect{\theta}^* = \argmin_{\vect{\theta}} & \multicolumn{3}{l}{\sum_{i=0}^{N}\ \  (\lambda_d \Gamma_{D,i} + \lambda_n \Gamma_{N,i}) + \sum_{j=5}^{J}\ \lambda_p \Gamma_{P,j}}\\
        \mathrm{s.t.} &  0 \leq \vect{\theta} \leq 2\pi\\
    \end{array}
    \label{eq:optNew}
\end{equation}

The optimization with these surprisingly simple modifications worked very well for us, scaling to rigs possessing north of fifty degrees of freedom, showing some robustness to starting position of the root, and retaining the benefits of cheap computation cost and full differentiability. We formulate and solve Eq.~\ref{eq:optNew} using the NLOPT optimization library \cite{johnson2017nlopt}.

\begin{figure*}
\centering
\includegraphics[trim={0 8cm 0 7cm},clip,width=1.0\linewidth]{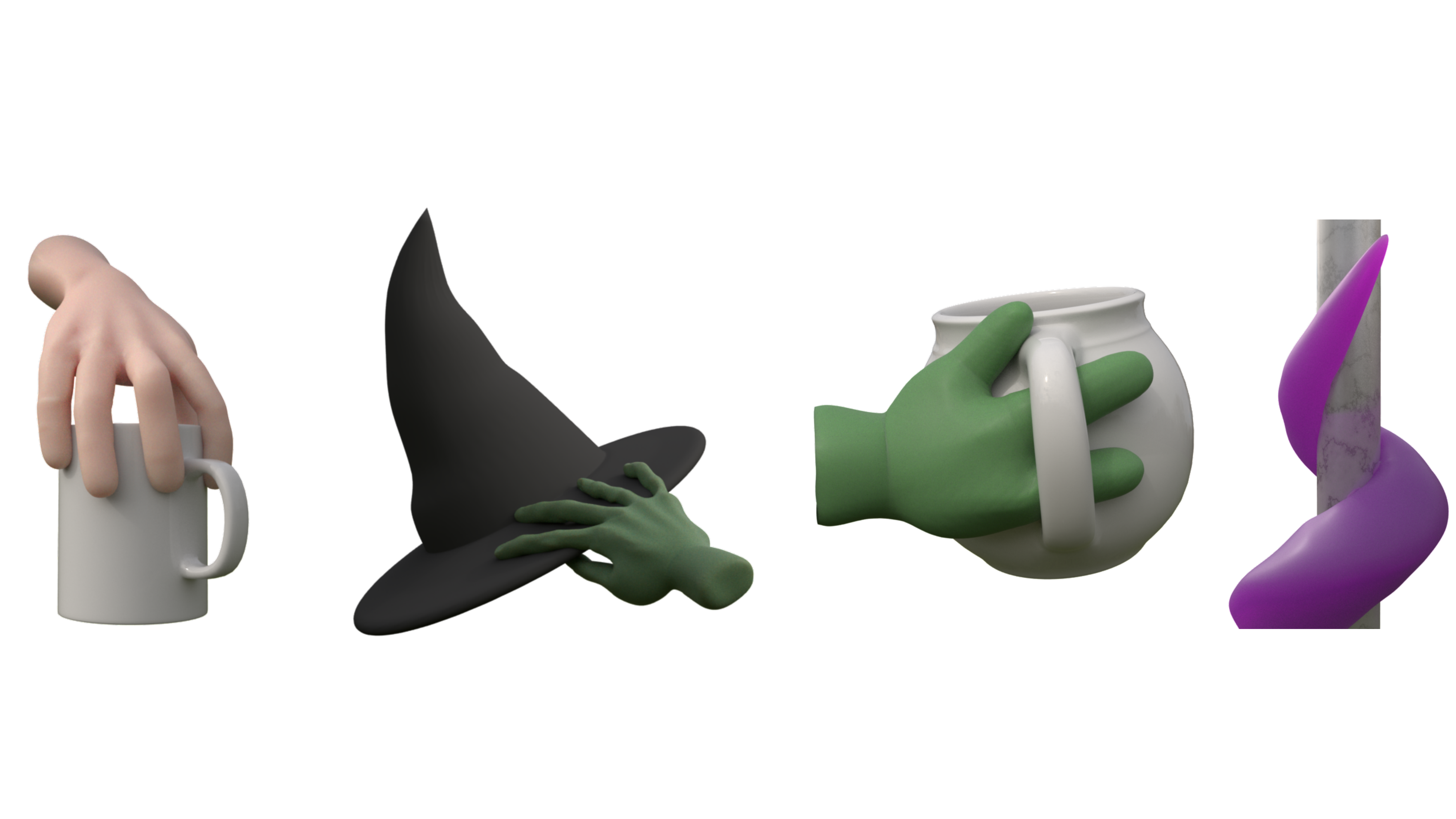}
\caption{Poses created using our contact edit framework across kinematically and morphologically diverse manipulators in realistic in-situ contexts.}
\label{fig:graspresults}
\end{figure*}

\section{Applications, Experiments, and Results}

We illustrate the benefits of our approach using a variety of objects and 4 custom manipulators: a human hand, a long-fingered witch hand, a three-fingered alien hand, and a tentacle. All experiments were conducted using our Maya plugin coupled with the geometry-central library \cite{geometrycentral}. 

\subsection{Static Grasps}

\begin{figure}
\centering
\includegraphics[trim={0 6cm 0 3cm},clip,width=1.0\linewidth]{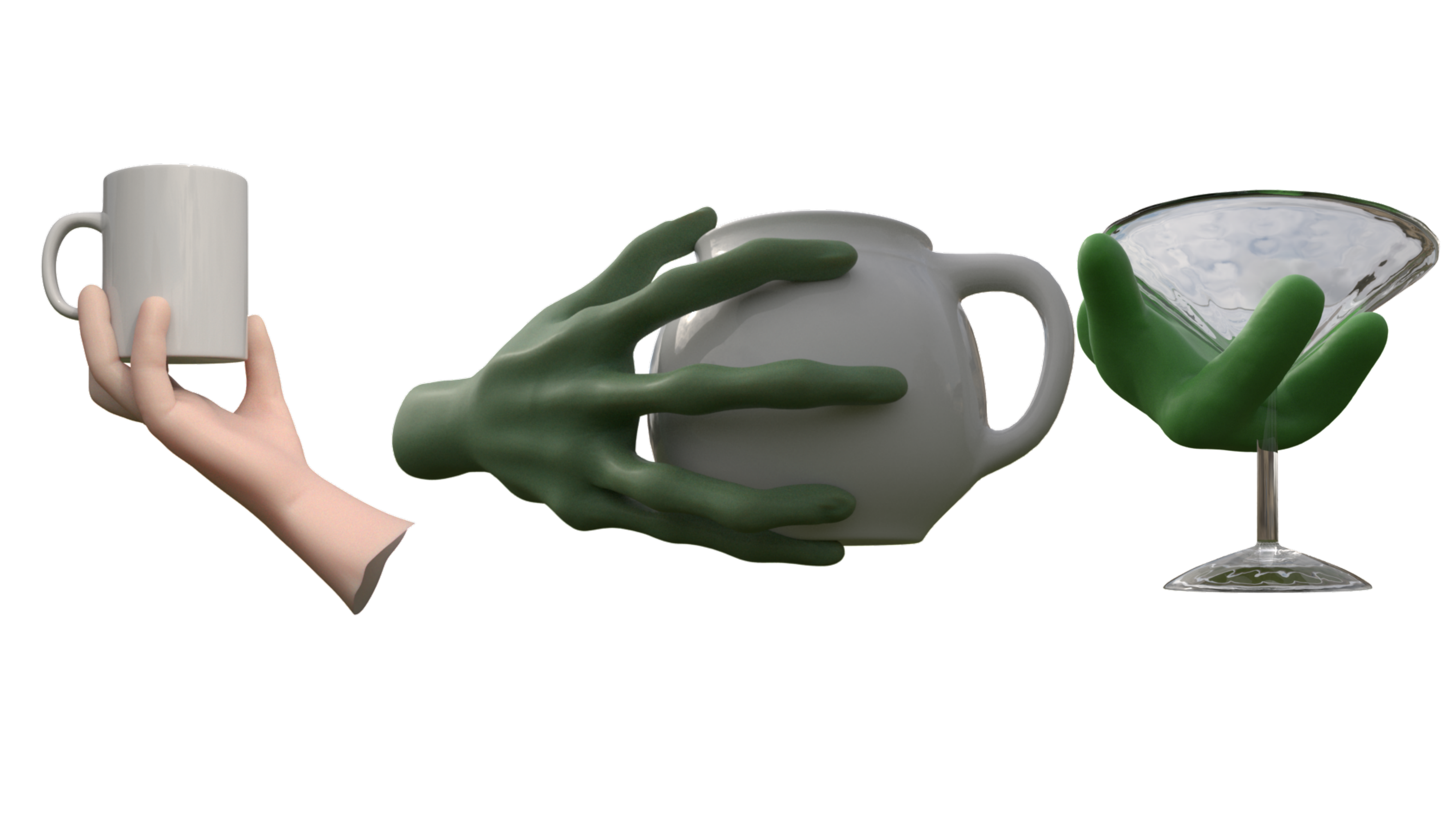}
\caption{Additional poses generated by our solver across various manipulators and geometries.}
\label{fig:additionalgraspresults}
\end{figure}

Figures \ref{fig:graspresults} and \ref{fig:additionalgraspresults} illustrate some high quality poses synthesized using our framework. Generating poses such as these using a traditional process is time consuming and iterative; however, our approach allows the user to quickly set up the desired shot by specifying contact locations, after which the corresponding manipulator pose is automatically calculated in a manner that accounts for both rig kinematics and skin deformation.

\subsection{Rapid Grasp Transitions}

\begin{figure}
\centering
\includegraphics[trim={0 6cm 0 3cm},clip,width=1.0\linewidth]{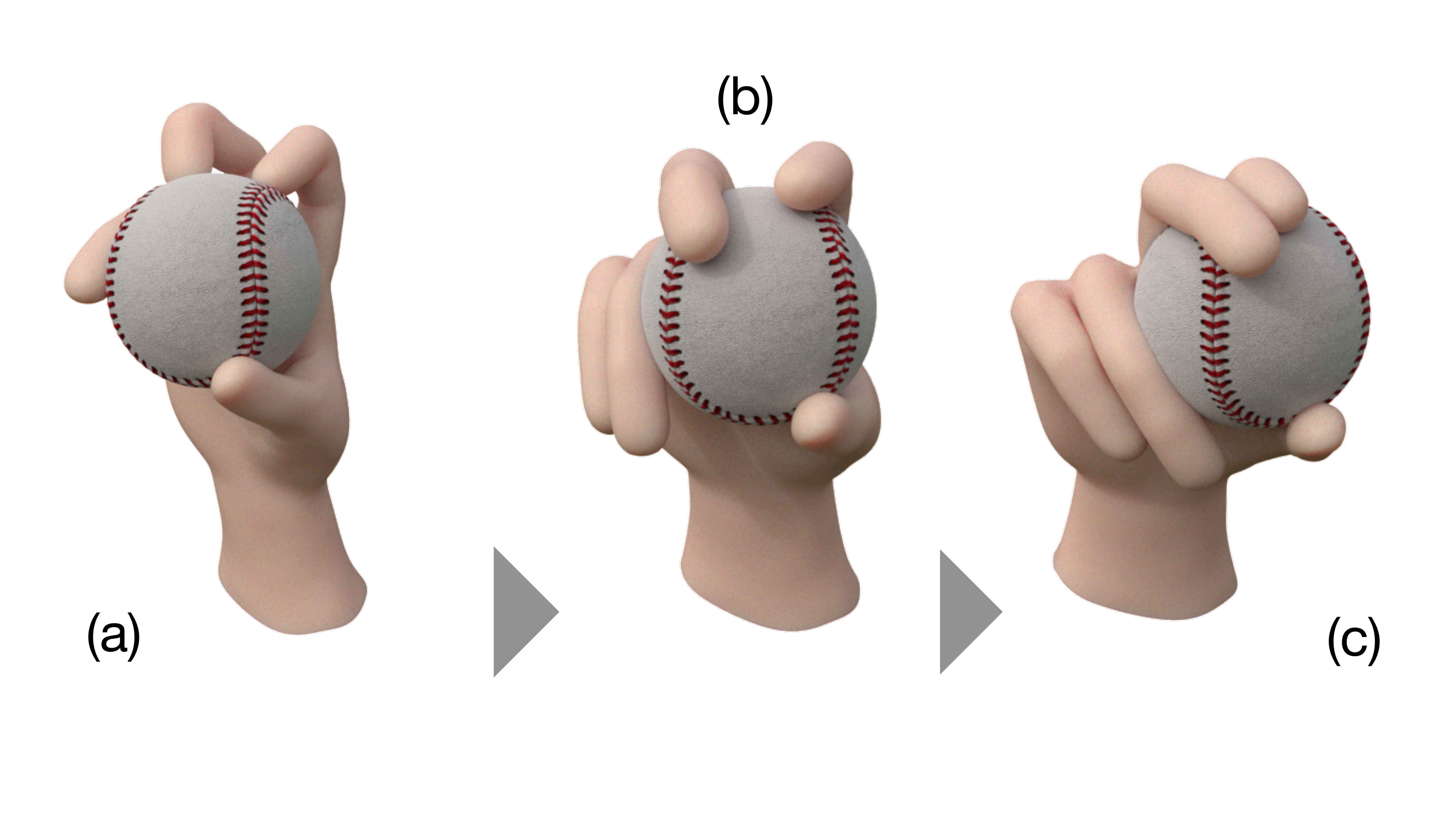}
\caption{Our technique can create smooth transitions between multiple baseball pitching grips, including a (a) knuckleball, (b) two-seam fastball, and (c) curveball.}
\label{fig:grasptransition}
\end{figure}

Figure \ref{fig:grasptransition} illustrates a sequence of transitions between multiple pitching grips, each of which was constructed through direct manipulation of contacts on the surface of a baseball. Each transition required exactly three manipulation operations: two translations and one rotation to go from the knuckleball to the fastball, and three translations to move from the fastball to the curveball. The entire sequence was generated by an untrained animator using reference photos in less than an hour after initial scene setup. Subtle details, such as closing the gaps between fingers, were made by creating corresponding contact patches on different fingers of the same mesh. The full sequence can be viewed in the supplementary video.

\subsection{Hierarchical Contact Transfer}

\begin{figure}
\centering
\includegraphics[trim={0 2cm 0 0},clip,width=\linewidth]{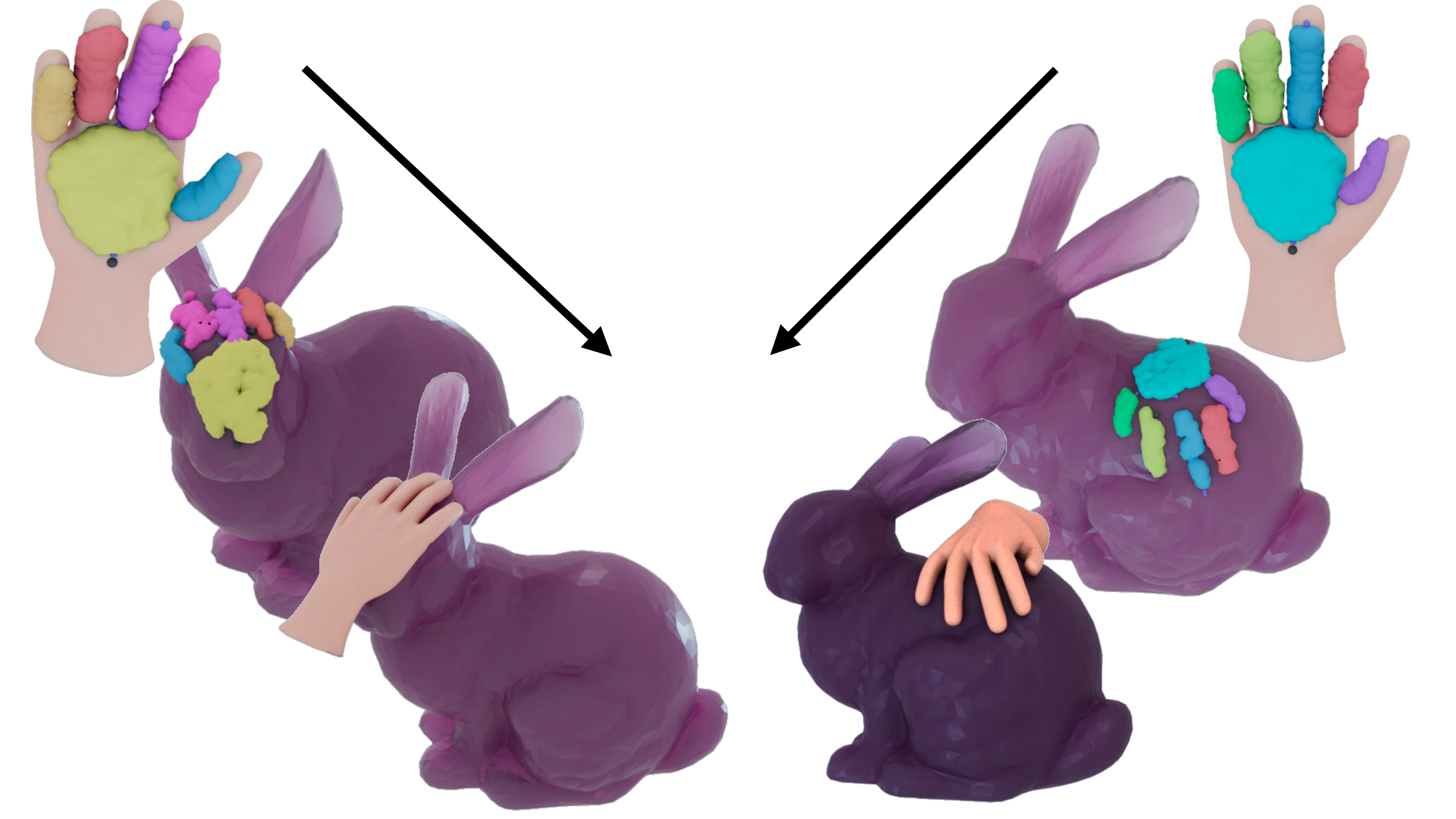}
\caption{Bulk transfer of a hierarchically organized identical set of contacts on the hand complies to local geometric features on the head (left) and body (right) of the Stanford bunny.}
\label{fig:bunnyregion}
\end{figure}

Artists can also paint multiple contacts onto the hand and then map them to an object in one broad sweep using hierarchical contact patch transfer. Figure \ref{fig:bunnyregion} shows an example. Contacts on the hand are mapped to different areas of the bunny with a few mouse clicks. Our as-isometric-as-possible transfer process adapts well to the different geometric features on the bunny surface.

\begin{figure}
\centering
\includegraphics[width=1.0\linewidth]{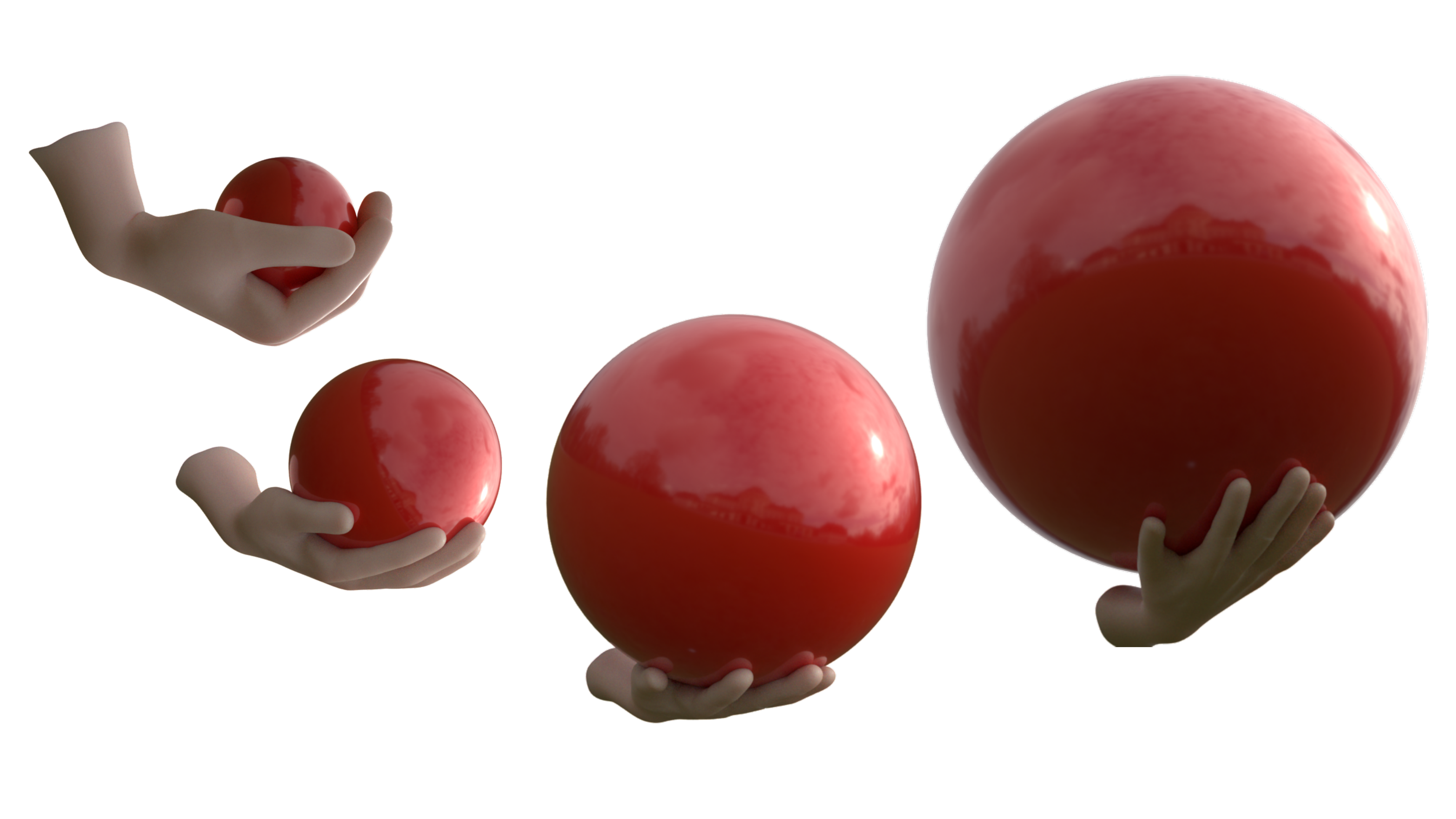}
\caption{Demonstration of a manipulator adapting its grip in response to variance in object scale.}
\label{fig:sphereexpansion}
\end{figure}

Figure \ref{fig:sphereexpansion} illustrates another example of this type of workflow, this time showing adaptation to changes in object scale -- a unique benefit to our approach. This sequence was constructed using 4 keyframes in under 45 minutes by an inexperienced animator. To do so, we transferred the contacts from the prior experiment to an identical starting point and axial direction of the sphere at each scale, and use the solution from the previous keyframe as a prior for the subsequent frame to reduce optimization time. The animated result can be seen in the supplementary video. Note that while alternative techniques such as drawing from pre-compiled taxonomies (e.g. \cite{sanso1994automaticgrasping,li2007shape}) may yield similar results in the case of human hands, our method does not require any pre-compiled hand pose priors and makes no assumptions of hand kinematics -- artists could easily replicate the same procedure on a non-anthropomorphic hand, for example, where taxonomies may be impossible to compile.

\subsection{Difficult Grasps on Higher Genus Topologies}

\begin{figure}
\centering
\includegraphics[width=1.0\linewidth]{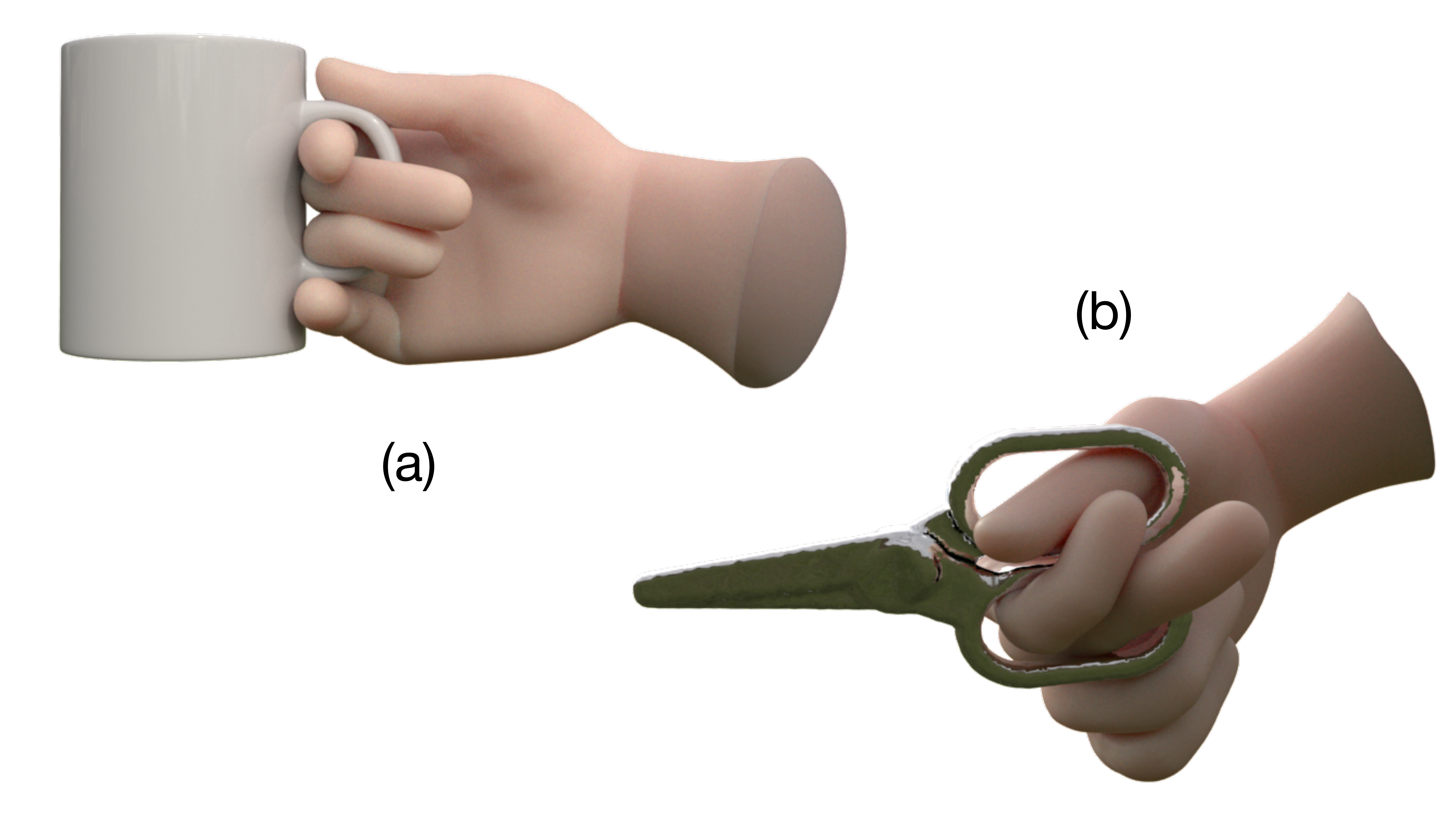}
\caption{Challenging grasps generated using our method (a) curling the fingers around the cluttered interior of a genus-1 mug handle and (b) between the two holes of a pair of genus-2 scissors.}
\label{fig:difficultgrasps}
\end{figure}

Routing multiple fingers through holes and around handles is an especially challenging task due to the large number of inter-object and self-intersections generated during traditional posing, many of which would require movement of the palm and a subsequent total repositioning of the fingers in the updated state to resolve. Figure \ref{fig:difficultgrasps} shows two examples. Using contacts allows the animator to designate which parts of the handles are desirable for contact and automatically generate a pose for the palm optimized for the desired grasp. Different fingers can also be routed towards specific handle regions to reduce self-intersections, while automated computation of the palm location can determine the optimal base pose position required to produce such a routing.

\subsection{Ablation Studies}

\begin{figure}
\centering
\includegraphics[width=1.0\linewidth]{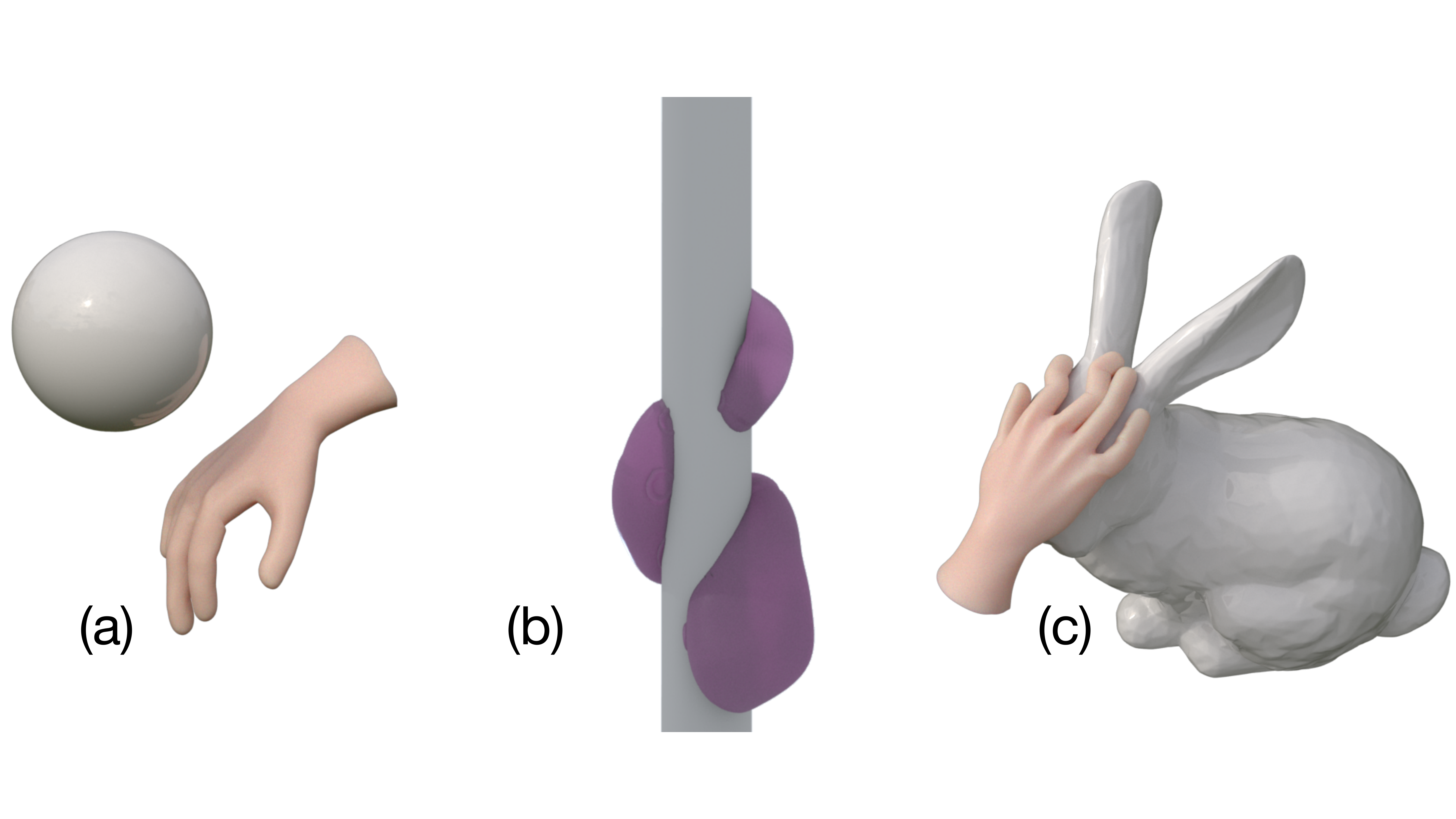}
\caption{Catastrophic failures resulting from the removal of optimization terms, including (a) distance, (b) normal, and (c) prior penalties.}
\label{fig:ablationstudy}
\end{figure}

To demonstrate the necessity of each term in Eq. \ref{eq:optNew}, we performed an ablation study in which each optimization term is individually removed. Figure \ref{fig:ablationstudy} illustrates the results of the study. As anticipated, removal of any terms results in catastrophic failure.

Removal of $\Gamma_D$ removes all incentive to progress towards the contacts, resulting in failure to even reach the desired object. Removal of $\Gamma_N$ permits negligence of the geometry at the contact point locations, which creates ambiguity in distinguishing the object interior and exterior. This omission caused the tentacle to find a solution which conformed to the inside of the pole during part of the wrap rather than the outside. Finally, removal of $\Gamma_P$ commonly results in highly undesirable pose artifacts such as contorted fingers and large regions of self-intersection. We note that this omission is particularly problematic in the context of unconstrained problems where there is otherwise no quantitative regularization to prevent unnatural behaviors from emerging.

\subsection{Extensions}

\begin{figure}
\centering
\includegraphics[trim={0 6cm 0 9cm},clip,width=1.0\linewidth]{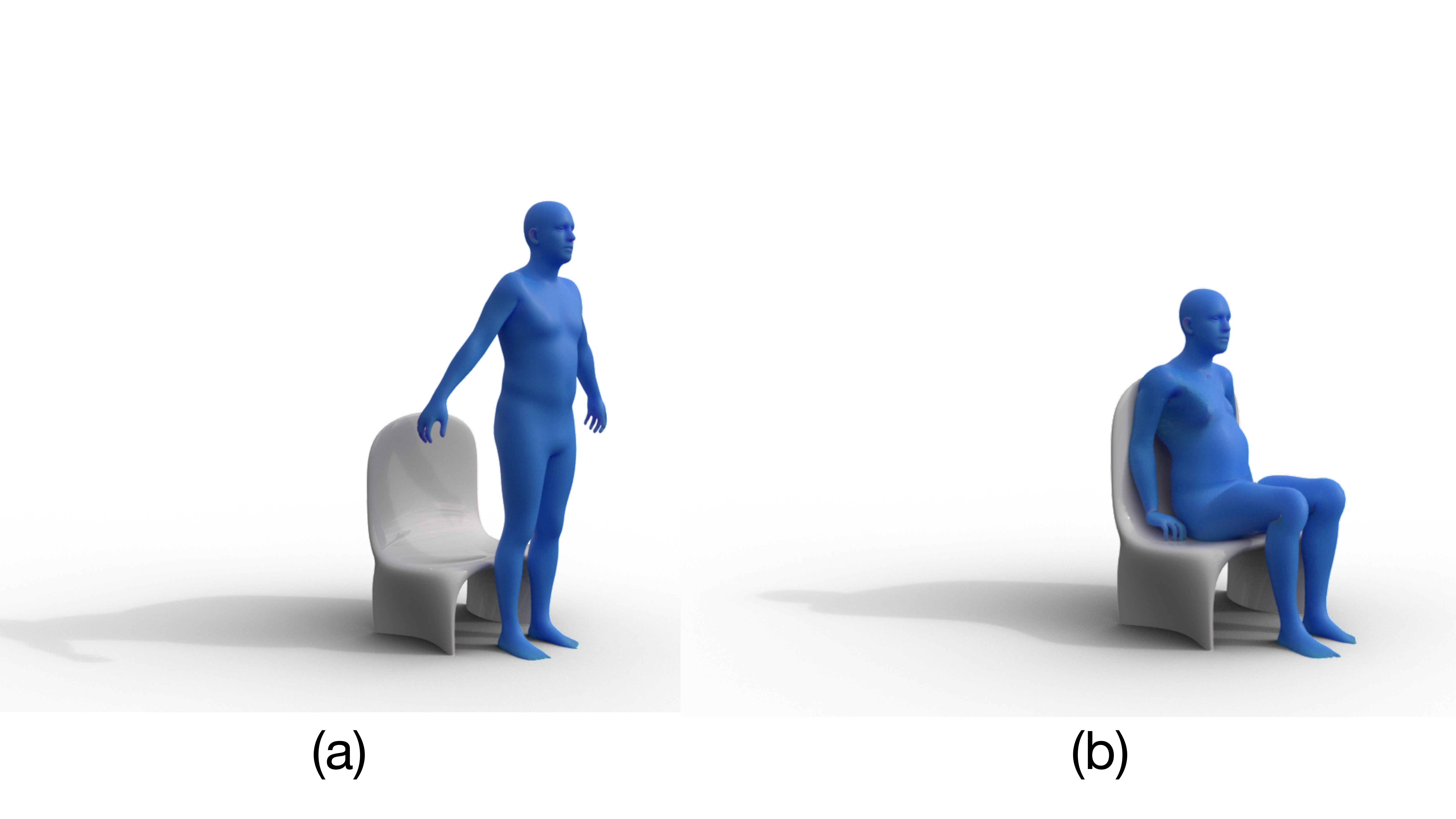}
\caption{Result of applying our method to designate contacts on a full body (a) in preparation for sitting in a chair and (b) settled into the final pose.}
\label{fig:fullbody}
\end{figure}

Although we have primarily considered manipulator-object interaction in this paper, the general-purpose nature of our contact formulation permits application to arbitrary contact-rich contexts.

Figure \ref{fig:fullbody} illustrates an example in which various parts of a full body are designated to make contact with the chair and floor. To animate this sequence from the initial standing position after specifying the contact sets, we start by fixing all joints and solving for the position of the root joint at the center of the pelvis. Next, we unlock the lower body and solve for the leg positions which align the feet with the floor. Finally, we unlock the left and right arm subsystems and solve in order to bring the arms inward. Another use case is to minimize foot movement during the in-betweens of the sitting procedure, which we can do by solving for the foot contacts while the remainder of the upper body continues with the sitting interpolation.

\section{User Studies}
\label{sec:UserStudies}

We conducted a small user study to gauge the effectiveness of our plugin in the hands of real animators.

\subsection{Subject Recruitment and Distribution}

Five subjects over age 18 were recruited from a combination of channels including internal university mailing lists, in-class announcements, and referrals. Subjects were expected to have basic working proficiency in Maya, including the ability to control rigs and change camera views. Three out of five subjects came exclusively from an art background, while two had some technical background.

\subsection{Setup}

\begin{figure}
\centering
\includegraphics[width=0.7\linewidth]{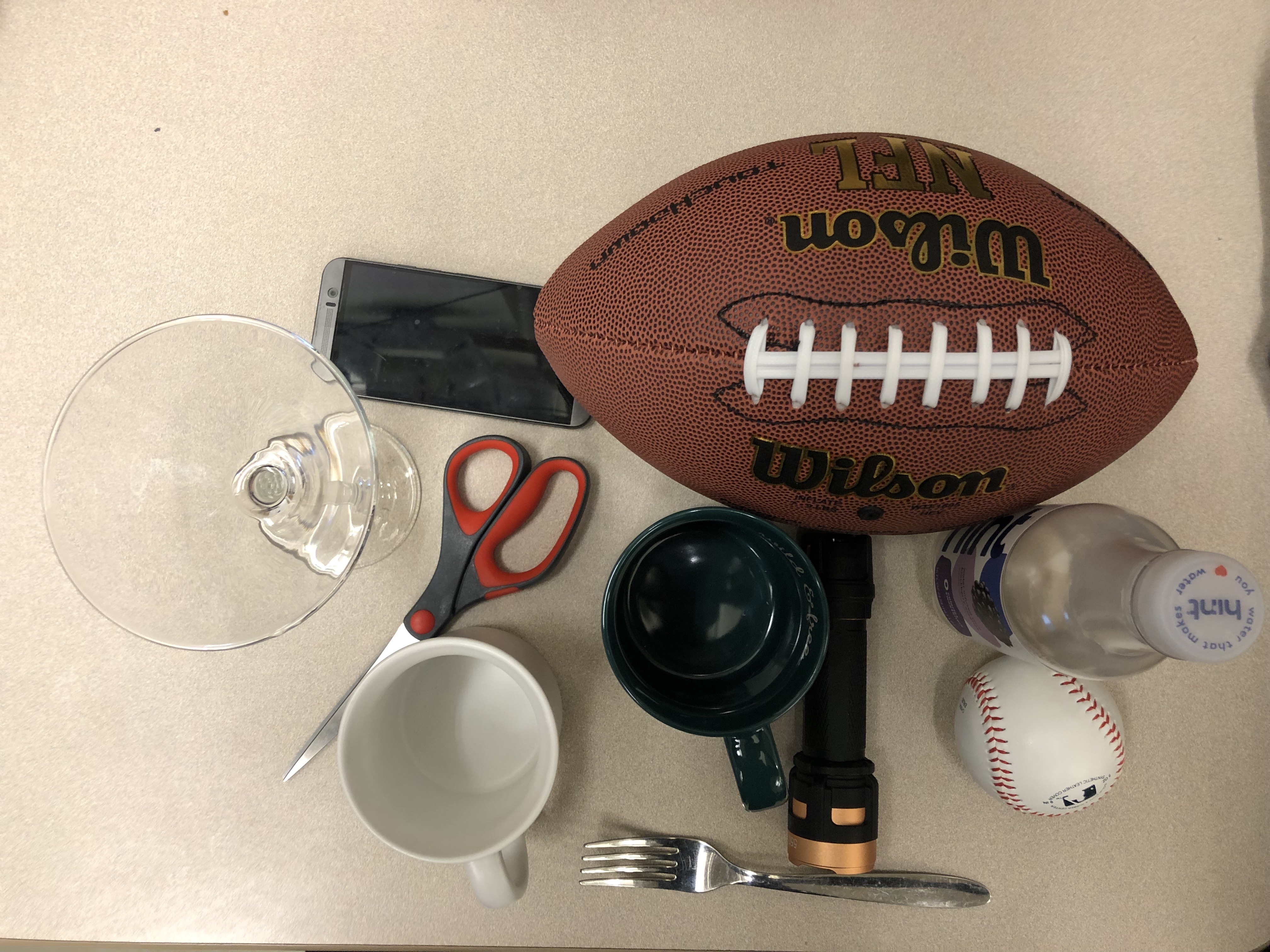}
\caption{All user study items}
\label{fig:userstudyitems}
\end{figure}

Participants were asked to partake in three one hour sessions and were compensated with a \$20 Amazon gift card at the end of each session for a total of \$60 per subject. Participants were asked to watch a 9 minute tutorial video prior to attending the first session but were not informed of session details in advance. All studies were conducted in-person with a proctor present during each session. Each session was conducted on a single workstation running an Intel i7 3.1 GHz processor with 8 GB of RAM. Sessions were screen-recorded without audio to limit exposure of personally identifiable information (PII). Participants were additionally provided a Maya scene with a skinned and rigged hand asset, a 3D digital scan of a physical object, and the physical object itself as reference material. 3D scans including geometry and textures were collected using an Artec Space Spider\footnote{https://www.artec3d.com/portable-3d-scanners/artec-spider}. Participants were permitted full internet use during each session and were permitted to ask unlimited questions. Figure \ref{fig:userstudyitems} includes all objects used during the studies. For simplicity, and to reduce the learning burden, we provided subjects access only to the default vertex selection paintbrush already included in Maya due to the high polygon counts of the 3D scanned meshes.

In the first session, participants were given an object by the proctor and asked to create as many good quality, natural looking poses as possible using the provided hand rig and object mesh. Participants were then asked to recreate the same poses on a best effort basis in the second session using view-only replicas of their original poses. Participants were tasked with performing the first and second sessions either using our plugin or existing Maya tooling only. The order of tool usage was randomized to reduce ordering bias. In the third session, participants repeated the first session procedure but were allowed to select any object in Figure \ref{fig:userstudyitems} and use our tool as much or as little as desired. In an effort to simulate a production workflow, participants were not allowed to modify the provided hand rig, skinning weights, object or hand geometry, or asset scales; however, subjects were allowed to perform rigid transformations to the object or rotate any of the hand joints. Each subject was played an audio recording of directions at the start of each session. Surveys were sent to subjects at the completion of each session.

Subjects were given access to the tutorial video and a textual reference during each session in which our plugin was used. Although participants were told to create as many high quality poses as possible or replicate as many poses as possible in the previous session using the plugin, our primary intention in the first plugin session was to provide subjects with hands-on experience with our tool, and in particular the processes of creating, manipulating, and clearing contacts, using and observing the optimization process, and navigating the general UI layout. To test the intuitiveness of plugin usage, we further randomly subdivided the first plugin session into two pools: participants in the first pool were provided guided suggestions by the proctor for exploiting plugin features, while those in the second pool only received answers to queries.

Rather than requiring users to designate an axis during contact creation time, we provided users with an automatically computed initial default axis to use with any newly created contact patch. We tried two methods for computing a default axis. In the first method, we simply connect the two furthest points on the patch with a geodesic. In the second, we use an approximation of the patch's medial axis as the default axis: we use the heat method \cite{crane2017shm} to compute a smoothed distance field $u$ from the boundary of the patch, compute the gradient $\nabla u$, and record points where $\nabla u$ changes direction by an angle of more than $\pi/2$ as points on the (smoothed) medial axis. These points were then greedily connected into a curve using Dijkstra's algorithm.

\subsection{Results}

Our findings were exciting and in some cases surprising. Although the subject pool was small, the response was generally positive. We elaborate upon detailed observations, positive feedback, and criticisms in the proceeding paragraphs.

One subject opted to use our plugin as a ``last-mile" solver by painting contacts after creating an initial pose manually, while the remainder, as expected, used our plugin as a ``coarse" solver by instead painting contacts in the beginning, solving for the pose automatically, and making minor adjustments manually at the end, if at all. 100\% of subjects opted to paint contacts on the hand first and subsequently transferred them to the object. Three subjects opted to change the auto-generated default axis to an orientation consistent across all contacts. The remaining two, who opted to keep the default axis in all cases, opted to bring the object close to the hand before transferring contacts. All subjects could easily tell when transfer results were undesired and were able to quickly alter orientations via contact rotation. Subjects who heavily used contact rotation and translation did so with fine-grained movements and expressed satisfaction with the result. Local deformation was used sparingly. Three subjects opted to start from scratch for each new pose, while two chose to return to a template contact distribution for re-use across multiple poses and expressed interest in having an existing library for common regions such as fingertips or upper palm regions. All subjects that made fine-grained adjustments post optimization did so through manual rig adjustment rather than creating or modifying contacts. With the exception of one subject who created very sparse contact sets, users never modified any weighting hyperparameters. All subjects opted to visualize the optimization progress to understand how the final solution was generated and, more interestingly, how they could modify variables such as object or contact placement to produce even more desirable results.

Despite the large variance in usage across even a limited subject pool, the response from all subjects in the third session was positive. One subject opted to use the plugin exclusively during the third session, and the remainder chose to use both toolsets together. 100\% of subjects claimed the plugin would be useful to other animators and found the tutorial video alone sufficient for using the tool regardless of whether guided suggestions were provided in the introductory session. The response to optimizer-generated poses was generally positive -- most subjects were very satisfied, while some opted to make minor changes after completion.

Dissatisfaction and criticisms were consistently associated with one of three issues: solutions to difficult optimization problems such as those which involved very large wrist movement, contact transfers to object locations with vastly different curvature than the source region, or generic UI layout and the lack of hotkeys. Subjects who understood optimization limitations deliberately tried to push the limits of the solver, while those who encountered some unwanted artifacts usually just modified the object transform slightly and re-solved. We also found that neither of our default axis generation methods were really successful -- regardless of the default axis provided, subjects either preferred to create their own axis or opted to move forward in spite of, rather than aided by, the generated default axis.

While the subject pool was small, these results do suggest that our tool is intuitive and predictable, and more importantly that those results held true across users with different existing skill, users with limited technical background, and in a variety of usage patterns. We found it especially encouraging that many subjects approached us directly after the final session and asked when the plugin would be available publicly, and even moreso that the inquires came from heavy Maya users. A number of subject-generated poses, along with more detailed results comparing usage time split per subject between our plugin and existing tooling, are available in the appendix.

\section{Qualitative Assessments}
\label{sec:QualitativeStudies}

To assess the quality of user study results, we additionally conducted a round of qualitative assessments.

\subsection{Subject Recruitment and Distribution}

We published and distributed a survey online to a total of 65 individuals with a background or interest in animation via a combination of internal mailing lists, individual outreach, and class announcements. We also solicited the individual response of an expert animator with ten years of experience. Subjects were required to be over age 18. All subject information, with the exception of the expert animator, was kept anonymous.

\subsection{Setup}

We collected 12 poses generated by users from the previous study which used our plugin in a substantial capacity and 12 poses created using fully manual FK tooling. Poses were selected based on overall quality by an internal review process. All selected poses are available in the appendix.

Each pose generated using the plugin was paired with another using only manual tooling primarily by user study subject ID, secondarily by item, and lastly at random with results from other subject IDs. All pairs of corresponding images are available in the appendix. Importantly, we note that the decision to primarily group by subject ID rather than object was to calibrate for individual skill level, both with respect to existing experience and the level of familiarity with our plugin vs. existing tooling. Since poses using the plugin were largely sourced from the third session (e.g. where subjects were allowed to choose their object), nearly all pairs present non-corresponding items. Our results therefore assess the general ability of an individual subject with and without access to our plugin, which we consider more meaningful than comparing results between subjects.

Survey subjects were presented with 12 pairs of images in A/B format spanning each of the following categories:

\begin{itemize}
    \item Overall visual appeal
    \item Realism / physical plausibility
    \item Complexity
    \item Natural appearance
\end{itemize}

\noindent Answer choice and category order was fully randomized to reduce ordering bias. Subjects were required to provide a response to all questions in order to receive credit for completion.

\subsection{Results}

Table \ref{table:qualityresults} tabulates the results of the study, where each answer choice is awarded a score of +1 for a win, 0 for a loss, and 0.5 for a tie. Selections were required to have a vote differential of 4 or more to be considered the majority choice.

\bigskip

\begin{table}[h!]
\centering
\caption{Qualitative study results}
\begin{tabular}{ |>{\centering\arraybackslash}M{2.5cm}|M{2cm}|M{2cm}| }
    \hline
    \multicolumn{3}{|c|}{Qualitative Study Results} \\
    \hline
     & Plugin Score & Manual Score \\
    \hline
    Overvall Visual Appeal & 8 & 4 \\
    \hline
    Physical Plausibility & 6 & 6 \\
    \hline
    Complexity & 4 & 8 \\
    \hline
    Natural Appearance & 6.5 & 5.5 \\
    \hline
\end{tabular}
\label{table:qualityresults}
\end{table}

\bigskip \medskip

The expert opinion was relatively consistent with aggregate results. Specifically, 9/12 plugin poses were preferred for visual appeal, 8/12 for physical plausibility, 4/12 for complexity, and 10/12 for natural appearance. Overall, poses generated by the plugin performed as well, if not better, than \textit{fully manual} posing in a majority of categories. Importantly, we note that the loss in complexity is expected since users were exposed to the plugin for the first time during the previous sessions. We note that poses such as those in Figure \ref{fig:difficultgrasps} are possible to generate, but require more experience. However, performance in the remaining categories clearly indicate that our tools have a substantial impact on pose quality.

\section{Comparisons}

\subsection{Contact Model Comparisons}
\label{sec:contactModelComparisons}

\begin{figure}
\centering
\includegraphics[width=1.0\linewidth]{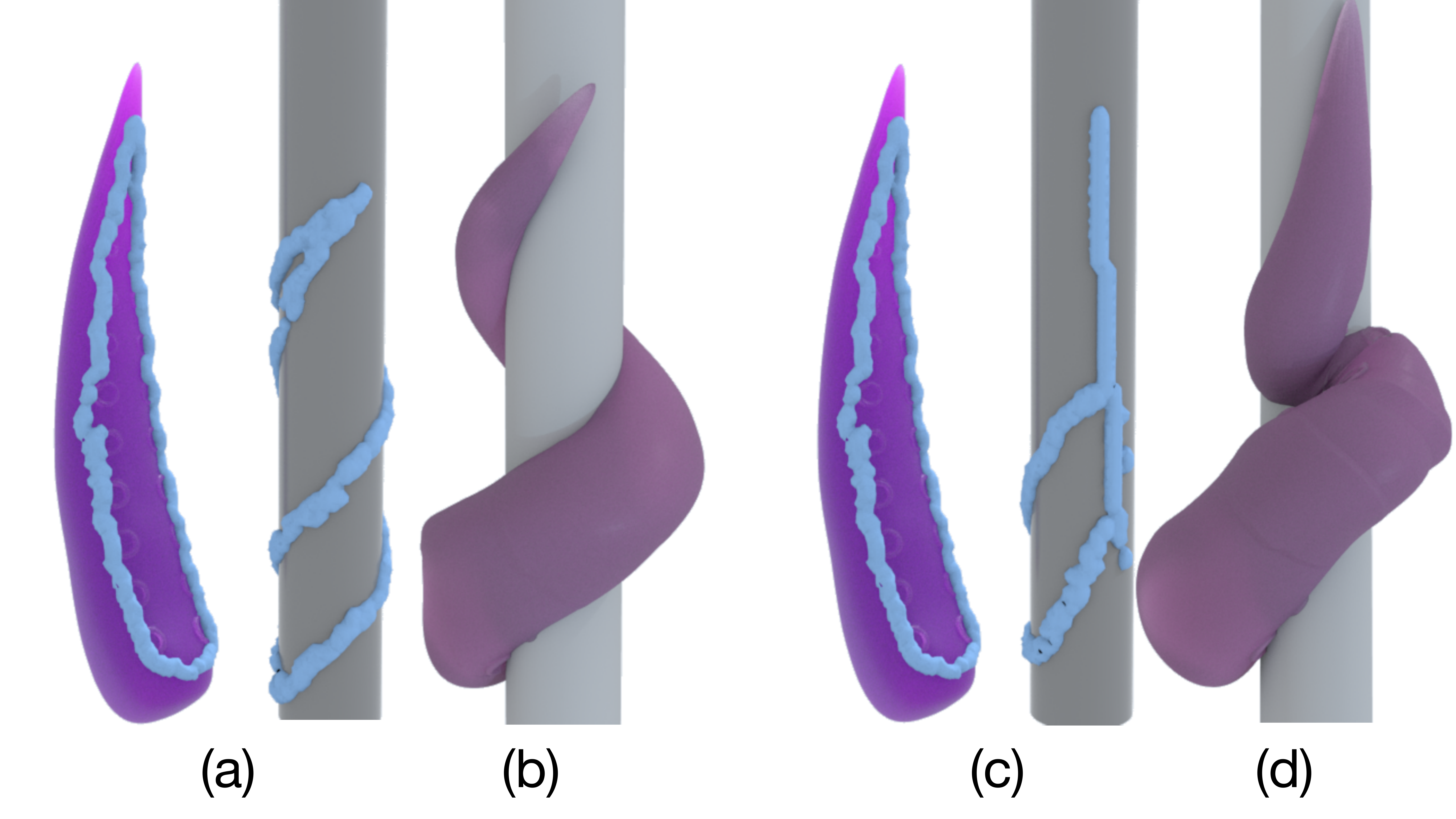}
\caption{Contact model behavior during transfer of a tentacle contact to a cylinder using (a) Our contact model and (b) its generated solution. (c) The contact model from \cite{lakshmipathy2022contacttransfer} and (d) its generated solution}
\label{fig:logmapcomparison}
\end{figure}

We compare our axis-based contact model to those of previous works \cite{lakshmipathy2021contacttracing,lakshmipathy2022contacttransfer}, which are boundary-based methods that store bounding contact points only and utilize a single-point log map embedding to respectively. Our model is optimized for usage as an artist tool, with the goal of minimizing the time taken both to digest an original user specification as well as to perform downstream editing and optimization.

It is possible to parameterize only the patch boundary and recover its interior via flood fill \cite{lakshmipathy2021contacttracing}; however, since our brush tools already enable easy user selection of areas, all points in the interior will be used for downstream optimization, and flood fill is slow to run on fine meshes, it is faster to keep the whole region rather than re-compute the interior per edit, especially since our method is fast enough to perform real-time editing operations without compromising robustness. Bypassing flood fill also enables our operations to be sampling-agnostic, which provides consistent editing behavior and performance across different triangulations.

The axis model is also significantly more robust than the single-point log map embedding \cite{lakshmipathy2022contacttransfer}, which notably suffers the same drawbacks of far-region distortion and foldover as prior works in the texture transfer literature. As illustrated in Figure \ref{fig:logmapcomparison}, these limitations can result in unexpected complete breakdowns of editing and transfer operations in perfectly reasonable use cases, ultimately leading to unpredictable behavior.

\subsection{Posing Method Comparisons}

\begin{figure}
\centering
\includegraphics[trim={0 4cm 0 4cm},clip,width=1.0\linewidth]{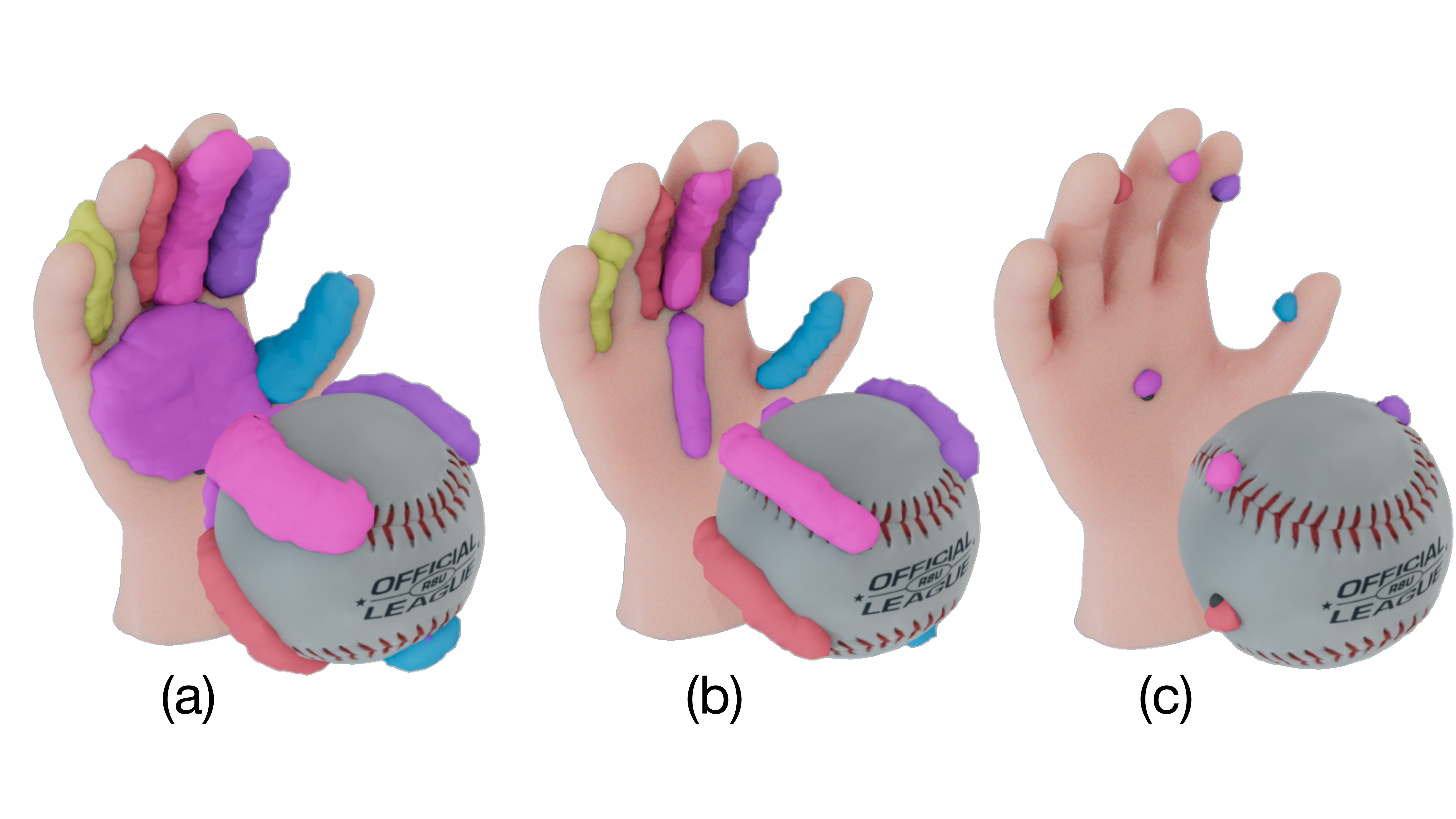}
\caption{Contact sets used as the basis for pose solution comparisons using (a) Our axis based area model (b) A curve-only model and (c) A single point model}
\label{fig:posecomparisoncontacts}
\end{figure}

\begin{figure}
\centering
\includegraphics[trim={0 4cm 0 2cm},clip,width=1.0\linewidth]{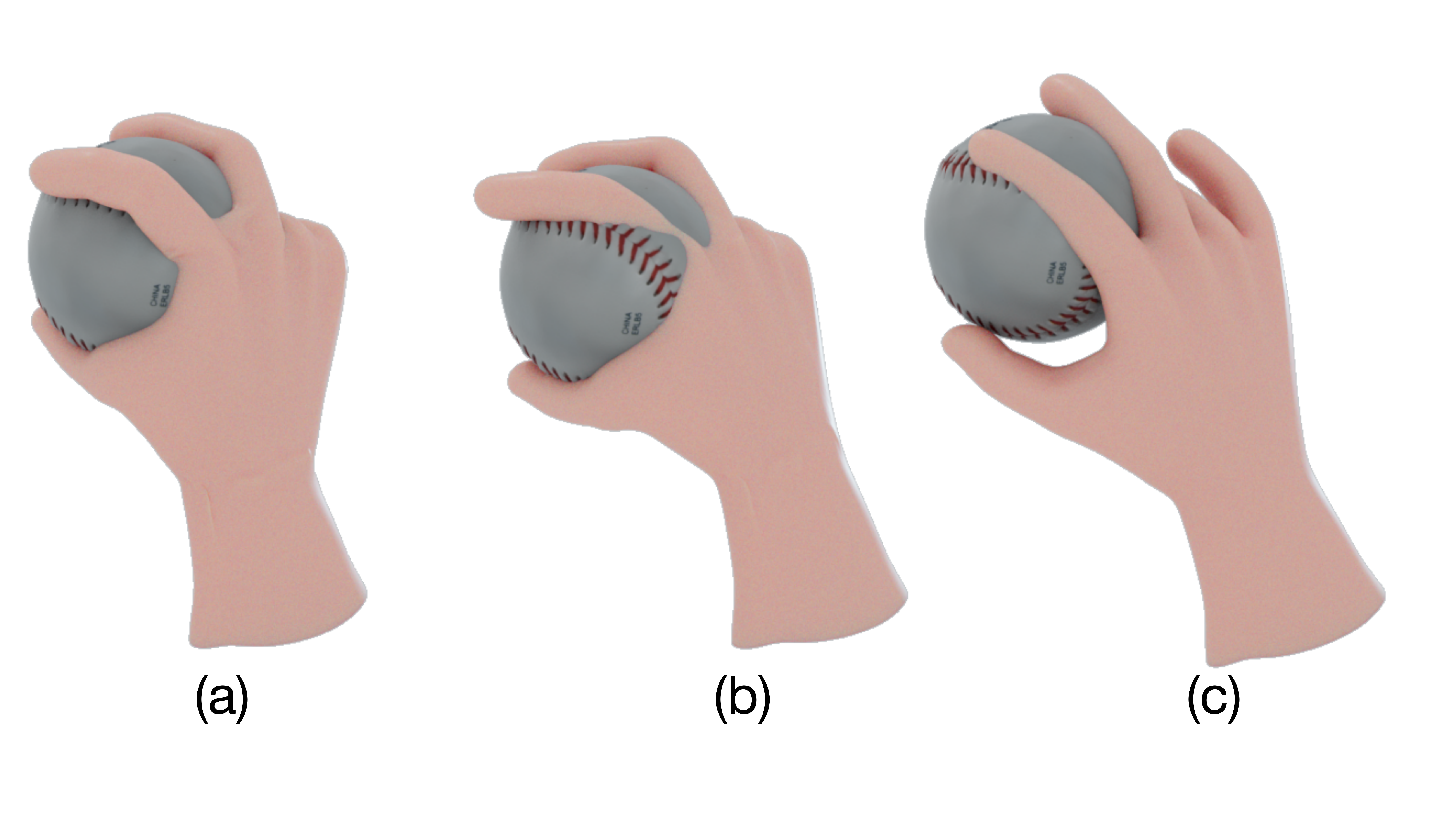}
\caption{Poses computed using (a) Our axis-based area model (b) A curve-only model and (c) A single point model}
\label{fig:posecomparison}
\end{figure}

We also perform a comparison of the solutions generated by our optimization framework utilizing contact areas against both traditional single point IK and a ``curve-only" variation inspired by the Line of Action \cite{guay2013lineofaction}. Figure \ref{fig:posecomparisoncontacts} illustrates the contact sets used as the basis for each solution. Full contact areas were approximated as single points and curves respectively on a best-effort basis. We found that responses were fairly consistent regardless of curve or point placement in ambiguous cases such as the large palm contact. Corresponding object points were selected individually in the single point formulation and were transferred using our axis-based solution in the case of the curve and area formulation, where the curve itself was used as the axis in the former case. All solutions were computed from same start pose using the same solver and number of iterations, but note that weighting coefficients $\lambda_d$, $\lambda_n$, and $\lambda_p$ had to be altered to account for differences in point density. Specifically, the values used are:

\bigskip \bigskip

\begin{table}[h!]
\centering
\caption{$\lambda$ weighting values used to generate results in Figure~\ref{fig:posecomparison}.}
\begin{tabular}{ |>{\centering\arraybackslash}M{2cm}|M{1.2cm}|M{1.2cm}|M{1.2cm}| }
    \hline
    \multicolumn{4}{|c|}{$\lambda$ Weighting Values} \\
    \hline
     & $\lambda_d$ & $\lambda_n$ & $\lambda_p$ \\
    \hline
    Single Point & 3.5 & 6.1 & 10.0 \\
    \hline
    Curve Only & 1.0 & 1.2 & 10.0 \\
    \hline
    Full Area & 1.0 & 1.0 & 10.0 \\
    \hline
\end{tabular}
\label{table:lambdavalues}
\end{table}

\bigskip \medskip

Figure \ref{fig:posecomparison} illustrates the poses computed under each formulation, while Table \ref{table:lambdavalues} provides the weighting coefficients used the generate the results. We note that both the single point and curve-only formulations resulted in solutions which produced highly unnatural finger configurations or significant penetration between the palm and the object respectively. The single point formulation also required extensive testing of contact point placement and hyperparameter tuning even to produce a somewhat similar result and was highly sensitive to changes, while curve and area formulations were less sensitive to changes. Importantly, we note that the choice of $\lambda$-values for the area based contacts not only produced the result in Figure \ref{fig:posecomparison}, but also \textbf{every single result in the entire paper, and all but one result in the appendix}. The curve formulation also notably converged to similar solutions regardless of how the palm curve was oriented, assuming the number of points comprising the curve stayed the same.

We hypothesized and verified that the curve and area formulations produce similar solutions if contacts are confined to regions with an implicitly defined orientation (e.g. fingers only). The failure of the curve formulation to generalize to regions such as the palm, however, underscores a general limitation: given two curves of the same length, it is not possible to non-uniformly weight the importance of one curve over another without either artificially making one curve sparser than the other or introducing more hyperparameters to adjust the weighting. In contrast, an area formulation implicitly provides non-uniform weighting capability by design. We note this observation to be particularly valuable in the context of artist tools since users already make this designation during creation of the contact areas themselves, which improves overall intuitive understanding of tool usage in comparison to tweaking weighting hyperparameters. Additionally, the results demonstrate that, more generally, an abundance of correspondences is valuable in addressing several well-known drawbacks in IK solvers.

\section{Drawbacks and Limitations}
\label{sec:DrawbacksAndLimitations}

\begin{figure}
\centering
\includegraphics{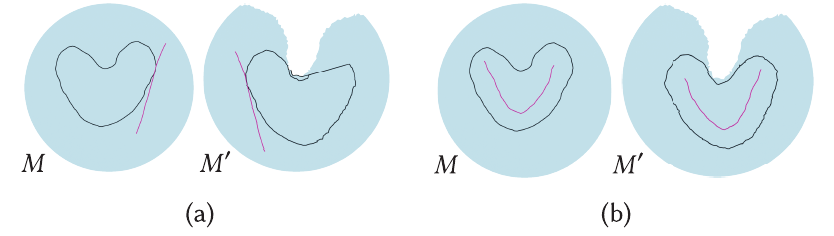}
\caption{(a) Our transfer method can produce large unwanted distortion if the axis (magenta) is drawn such that the patch parameterization depends on geodesics expected to travel outside of a non-convex domain. (b) However, desired results can still be obtained by simply choosing the axis to lie entirely within the patch region.}
\label{fig:transfer_failure_example}
\end{figure}

\begin{figure}
\centering
\includegraphics[width=1.0\linewidth]{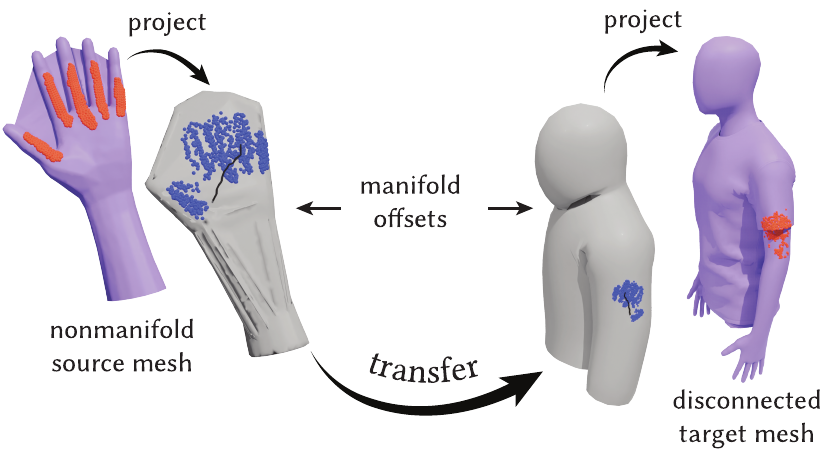}
\caption{Although the geodesic tracing step of our transfer process is not directly applicable to nonmanifold and disconnected surfaces, contact transfer can still be achieved between such surfaces via the use of a manifold offset mesh. Parameterization and transfer occurs between the offset surfaces, and patches are transferred between the original and offset surfaces via projection. Offset surfaces in this figure were constructed using the ``shrinkwrap'' modifier in the open-source 3D software Blender.}
\label{fig:manifoldoffset}
\end{figure}

Patch transfer can fail if there exist points $p$ on the patch and $q$ on the axis such that $d_{M}(p,q) \neq d_{M'}(p',q')$, i.e. the geodesic distance between $p$ and $q$ on $M$ does not equal the distance between their corresponding points $p'$, $q'$ on $M'$; Figure \ref{fig:transfer_failure_example}a illustrates such a failure. However, the user can draw an alternate axis that yields the expected results (Figure \ref{fig:transfer_failure_example}b).


Parameterization quality may also degrade for certain choices of axes, for example axes with too few control points to adequately parameterize a patch with a complex shape, or axes that deviate significantly from the general shape of the patch. In practice, our method produces reasonably robust transfers even for a highly non-intuitive choice of axis (inset).

\setlength{\columnsep}{1em}
\setlength{\intextsep}{0em}
\begin{wrapfigure}{r}{121pt}
\centering
\includegraphics{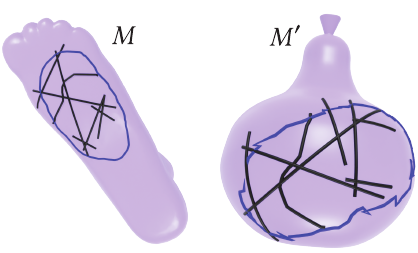}
\end{wrapfigure}

Our contact model is currently not well-suited for grasps that do not have sizeable areas of contact with the target object as well as grasps that span concavities or holes, such as the teeth of a comb or the holes of a button. While possible to model the contact as a set of many small patches, patch transfer and manipulation will be more susceptible to error. We also note that our optimization method, as illustrated by Figure~\ref{fig:posecomparison}(c), struggles when contact information is sparse, particularly in disambiguation of multiple solutions or resolution of uncanny artifacts.

Our method of contact parameterization and transfer is not designed for contact regions that belong to multiple disjoint surfaces, such as a figurine with separate layers of clothing. However, transferring such contacts is still possible under our framework. One option is to separately parameterize and transfer each portion of the contact region that belongs to a different surface. Alternatively, one can construct a single offset surface that encloses all components of either the source or target surface, project the contact region to/from the offset surface, and perform contact parameterization and transfer between the offset surfaces. On the other hand, transfer via a proxy surface often induces additional inaccuracy (see Figure~\ref{fig:manifoldoffset}). Robustly handling multi-component surfaces remains a challenge for future work.

Our solver produces locally optimal solutions only and tends to struggle with finding solutions which require large amounts of wrist movement. User study subjects expressed little concern with this limitation in practice since modifying the object transform or moving the wrist or object into a more favorable start position required minimal effort; however, increased sensitivity to initialization renders our approach more problematic in automation-focused settings.

Our current contact model also assumes manifold meshes, which is needed for the log map computations and tracing geodesics. Strictly speaking, however, we need only require that the mesh be locally manifold, rather than globally manifold. Specifically, each point on the axis on the source mesh needs to be manifold during contact parameterization, and the contact region on the target mesh must be manifold in order to reliably trace geodesics. Alternatively, one can perform mesh repair, or project onto/from a manifold offset mesh such as in Figure~\ref{fig:manifoldoffset}.

Other shortcomings involve the workflow of working with contacts. This issue was especially prevalent in non-human hands since the animator was expected to internally visualize the contact set without being able to replicate the pose themselves in the physical world. Maintaining finger spacing during transfer proved challenging as well, although hierarchical transfer partially combats the problem.

Two areas of future work are to reduce the time taken to draft contacts and extend our pose drafting capabilities to both full animations and VR / AR. Libraries of contact sets could assist with the former; however, full prediction is a notoriously challenging problem which requires a more extensive exploration into learning-based methods. The latter would likely require formulation of new interfaces, as our posing tools currently do not account for items such as interpolation, motion smoothness, or mid-air contact painting. We are excited by applications such extensions will enable.

\section{Conclusion}

In summary, we have presented EAD artist tools which enable intuitive modeling of high complexity, contact-rich manipulator-object interactions through the direct manipulation of contact areas. To do so we proposed a novel contact model formulation centered on a log map-based parameterization computed with respect to a user-chosen axis, which we demonstrate to be scalable to large contact regions, highly robust to the choice of axis and surface sampling variations, and capable of providing real-time, robust, and predictable updates in responses to edits including dragging, rotation, hierarchical composition, and transferring between surfaces. We have also introduced an optimization framework that is scalable to high DOF manipulators through two modifications of an existing formulation. Combined, our framework provides the first direct and responsive means of working with contact regions, which we demonstrate can lead to the synthesis of high quality results while allowing animators to retain full artistic control. We are confident that the foundations laid in this paper will serve as a strong basis for future studies and are hopeful that the combination of our distributed Maya plugin and emphasis on practicality will spur interest and encourage participation from the broader community in this line of work.

\begin{acks}
The authors would like to thank Helena Yang, Selena Sui, and Robin Zhang for their helpful contributions in the development of the plugin. We would also like to acknowledge all anonymous user study and survey subjects for their participation, and the anonymous reviewers for their helpful suggestions in revising our manuscript. We also thank Meta Platforms Inc. for access to the 3D scanner. This material is based upon work supported by the AI Research Institutes program supported by NSF and USDA-NIFA under AI Institute for Resilient Agriculture, Award No. 2021-67021-35329 and NSF award CMMI-1925130.
\end{acks}

\bibliographystyle{ACM-Reference-Format}
\bibliography{main}

\newpage 

\appendix

\noindent{\huge APPENDIX}

\section{Pseudocode}
\label{sec:pseudocode}

We assume the following atomic functions:
\begin{itemize}
\item \(\Proc{ExactGeodesic}(p, q)\) -- computes the exact discrete geodesic between two barycentric points $p$ and $q$ on the same triangle mesh $M$. Our plugin uses MMP.
\item \(\Proc{InitialDirection}(g)\) -- returns the initial direction of a geodesic $g$ in the tangent space of its origin point, expressed as a complex number.
\item \(\Proc{EndingDirection}(g)\) -- returns the ending direction of a geodesic $g$ in the tangent space of its ending point, expressed as a complex number.
\item \(\Proc{TraceGeodesic}(p,z)\) -- trace a geodesic starting from a barycentric point $p$, along a direction and distance specified by the complex number $z$. 
\item \(\Proc{ClosestPoints}(P, A)\) --  computes the closest point $a_j\in A$ to each point in $p_i\in P$, returning a vector $\Csf_P$ such that $\Csf_P[i]=j$. Our plugin uses the Vector Heat Method \cite{sharp2019vhm}.
\end{itemize}

\begin{algo}{\Proc{ParameterizeAxis}$(A)$}
\label{alg:ParameterizeAxis}
\begin{algorithmic}[1]
    \InputConditions{An array $A = [a_1,\dots,a_m]$ of barycentric points on some triangle mesh $M$.}
    \OutputConditions{A vector $\Phi_A\in\CC^{m}$ where $\Phi_A[i] = \phi_i$, the turning angle at $a_i$; a vector $\Lsf_A\in\RR^m$ where $\Lsf_A[i]$ is the geodesic distance between $a_i$ and $a_{i+1}$; and a vector $\Tsf_A\in\CC^m$ where $\Tsf_A[i]=t_i$ is the ``tangent direction'' at $a_i$.}
    \State {\(\Phi_A, \Lsf_A, \Tsf_A \gets 0^m\)} \Comment{initialize to length $m$}
    \State {\(s_i\gets 0\)} \Comment{track the direction of the incoming geodesic}
    \For {\(i = 1 \text{ to } m-1\)}
        \State {\(g_i \gets \Proc{ExactGeodesic}(a_i, a_{i+1})\)}
        \State {\(t_i \gets \Proc{InitialDirection}(g_i)\)} \Comment{tangent direction of $A$ at $a_i$}
        \State {\(s_i\gets \Proc{EndingDirection}(g_i)\)}
        \State {\(\Lsf_A[i] \gets \Proc{Length}(g_i)\)}
        \State {\(\Tsf_A[i] \gets t_i\)}
        \If {\(i > 1\)}
            \State {\LeftComment{Compute turning angle at interior points $a_i$}}
            \State {\(\Phi_A[i] \gets t_i/s_i\)} 
        \EndIf
    \EndFor
    \State {\LeftComment{Special case for defining the tangent direction at the endpoint.}}
    \State {\(\Tsf_A[m] \gets s_i\)}
    \State {\Return \(\Phi_A, \Lsf_A, \Tsf_A\)}
\end{algorithmic}
\end{algo}

\begin{algo}{\Proc{ParameterizePatch}$(P, A, \Tsf_A)$}
\label{alg:ParameterizePatch}
\begin{algorithmic}[1]
    \InputConditions{A set $P = \{p_1,\dots,p_n\}$ of barycentric points on some triangle mesh $M_1$; an array $A=[a_1,\dots,a_m]$ of barycentric points on the same mesh $M_1$; and a complex vector $\Tsf_A\in\CC^m$ representing the tangents of $A$.}
    \OutputConditions{A map $\Esf_{P}^A: P\to A\times\CC$ that maps each $p_i\in P$ to a tuple $(j, z_i)$ such that $a_j$ is the point on $A$ closest to $p_i$, $|z_i|$ is the geodesic distance between $p_i$ and $a_j$, and $\arg(z_i)$ is the direction at which a geodesic must leave $a_j$ to arrive at $p_i$, expressed in the tangent space of $a_j$.}
    \State {\LeftComment{Compute closest point on $A$ for each point in $P$}}
    \State {\(\Csf_P \gets \Proc{ClosestPoints}(P, A)\)}
    \For {\(p_i\in P\)}
    \State {\(j\gets \Csf_P[i]\)} \Comment{index of closest point}
    \State {\(a_j\gets A[j]\)} \Comment{get the closest point on $A$}
    \State {\(h_i \gets \Proc{ExactGeodesic}(a_j, p_i)\)}
    \State {\(r_i\gets \Proc{Length}(h_i)\)} \Comment{get distance from $a_j$ to $p_i$}
    \State {\LeftComment{Compute direction in tangent space of $a_j$}}
    \State {\(\theta_i \gets \Proc{InitialDirection}(h_i)\)}
    \State {\LeftComment{Compute direction relative to the tangent direction of $A$}}
    \State {\(\theta_i \gets \theta_i / \Tsf_A[j]\)}
    \State {\LeftComment{Encode distance \& direction as a complex number}}
    \State {\(z_i\gets r_ie^{i\theta_i}\)} 
    \State {\(\Esf_{P}^A[p_i] \gets (j, z_i)\)}
    \EndFor
    \State {\Return \(\Esf_{P}^A\)}
\end{algorithmic}
\end{algo}

\begin{algo}{\Proc{ReconstructAxis}$(\Phi_A, \Lsf_A, a_1', t'_1)$}
\label{alg:ReconstructAxis}
\begin{algorithmic}[1]
    \InputConditions{A vector $\Phi_A\in\CC^{m}$ where $\Phi_A[i] = \phi_i$, the turning angle at $a_i\in A$; a vector $\Lsf_A\in\RR^m$ containing the lengths of the geodesic segments comprising the same axis $A$; an initial barycentric point $a'_1$ on some target triangle mesh $M_2$; and an initial direction $t'_1$ as a unit-magnitude complex number.}
    \OutputConditions{An array $A' = [a'_1,\dots,a'_m]$ of barycentric points on $M_2$; and a vector $\Tsf_{A'}\in \CC^m$ whose $i$th element is the ``tangent direction'' at $a'_i$.}
    \State {\(A'\gets []\)} \Comment{initialize empty array}
    \State {\(\Tsf_{A'}\gets 0^m\)} \Comment{initialize empty vector}
    \State {\LeftComment{Place the first two points of $A'$.}}
    \State {\(\Proc{Append}(A', a'_1)\)} \Comment{add the first point of $A'$}
    \State {\(z\gets \Lsf_A[1]t'_1\)} \Comment{form distance \& direction for geodesic tracing}
    \State {\(g'_1\gets\Proc{TraceGeodesic}(a'_1, z)\)}
    \State {\(a'_2\gets\Proc{EndingPoint}(g'_1)\)} 
    \State {\(\Proc{Append}(A', a'_2)\)} \Comment{add the second point of $A'$}
    \State {\(\Tsf_{A'}[1]\gets t'_1\)} \Comment{record axis tangent}
    \State {\LeftComment{Record ending direction of $g'_1$ at $a_2$}}
    \State {\(s_i \gets \Proc{EndingDirection}(g'_1)\)} 
    \State {\LeftComment{Trace the rest of $A'$.}}
    \For {\(i = 2 \text{ to } m-1\)}
    \State {\LeftComment{Negate turning angle to get mirror image of $A$}}
    \State {\(\phi'_i \gets \overline{\Phi_A[i]}\)}
    \State {\LeftComment{Form distance \& direction for geodesic tracing}}
    \State {\(z\gets \Lsf_A[i] * s_i * \phi'_i\)} 
    \State {\(g'_i \gets \Proc{TraceGeodesic}(A'[i], z)\)} 
    \State {\(a'_{i+1}\gets \Proc{EndingPoint}(g'_i)\)} 
    \State {\(\Proc{Append}(A', a'_{i+1})\)} \Comment{add the next point of $A'$}
    \State {\(\Tsf_{A'}[i]\gets \Proc{InitialDirection}(g'_i)\)} \Comment{record axis tangent}
    \State {\LeftComment{Record ending direction of $g_i$ at $a_{i+1}$}}
    \State {\(s_i \gets \Proc{EndingDirection}(g'_i)\)} 
    \EndFor
    \State {\Return \(A', \Tsf_{A'}\)}
\end{algorithmic}
\end{algo}

\begin{algo}{\Proc{ReconstructPatch}$(A', \Tsf_{A'}, P, \Esf_{P}^A)$}
\label{alg:ReconstructPatch}
\begin{algorithmic}[1]
    \InputConditions{A set $P = \{p_1,\dots,p_n\}$ of barycentric points on some triangle mesh $M_1$, and its parameterization $\Esf_{P}^A$ with respect to an axis $A$; a transferred axis $A'$ corresponding to $A$, and its tangents $\Tsf_{A'}$ on some target triangle mesh $M_2$.}
    \OutputConditions{A set $P' = \{p'_1,\dots,p'_n\}$ of barycentric points on the target triangle mesh $M_2$.}
    \State {\(P' \gets \{\}\)} \Comment{initialize empty set}
    \For {\(p_i\in P\)}
    \State {\((j, z_i) \gets \Esf_{P}^A[p_i]\)} \Comment{get info from patch parameterization}
    \State {\(t_j \gets \Tsf_{A'}[j]\)} \Comment{get tangent direction at $a'_j$}
    \State {\(z_j \gets \overline{z_j}\)} \Comment{negate angle so $P'$ is the mirror image of $P$}
    \State {\(z \gets \overline{z_i}/t_j\)} \Comment{take direction relative to tangent direction of $A'$}
    \State {\(p'_i \gets \Proc{TraceGeodesic}(A'[j], z)\)} \Comment{trace geodesic on $M_2$}
    \State {\(\Proc{Insert}(P', p'_i)\)}  \Comment{record where the geodesic ended as $p'_i$}
    \EndFor
    \State {\Return \(P'\)}
\end{algorithmic}
\end{algo}

\begin{algo}{\Proc{TransferPatch}$(P, A, a'_1, t'_1)$}
\label{alg:TransferPatch}
\begin{algorithmic}[1]
    \InputConditions{A set $P = \{p_1,\dots,p_n\}$ of barycentric points on some triangle mesh $M_1$ representing the source patch; an array $A=[a_1,\dots,a_m]$ of barycentric points also on $M_1$ representing a user-drawn axis; an initial point $a'_1$ on some target mesh $M_2$; an initial direction $t'_1$ as a unit-magnitude complex number.}
    \OutputConditions{A set $P' = \{p'_1,\dots,p'_n\}$ of barycentric points on $M_2$, representing the mirror image of an approximately-isometric embedding of the source patch into $M_2$.}
    \State {\((\Phi_A, \Lsf_A, \Tsf_A) \gets \Proc{ParameterizeAxis}(A)\)}
    \State {\(\Esf^A_P \gets \Proc{ParameterizePatch}(P, A, \Tsf_A)\)}
    \State {\(A' \gets \Proc{ReconstructAxis}(\Phi_A, \Lsf_A, a'_1, t'_1)\)} \Comment{transfer axis to $M_2$}
    \State {\(P'\gets \Proc{ReconstructPatch}(A', \Tsf_{A'}, \Esf_{P}^A)\)} \Comment{transfer patch to $M_2$}
    \State {\Return $P'$}
\end{algorithmic}
\end{algo}

\section{Additional User Study Metrics}







\begin{table}[h!]
\centering
\caption{More detailed breakdowns of usage times of our plugin vs. existing tooling per subject}
\begin{tabular}{ |>{\centering\arraybackslash}M{2.5cm}|M{2cm}|M{2cm}|M{2cm} }
    \hline
    \multicolumn{3}{|c|}{Plugin Usage Results} \\
    \hline
    Subject \# & Usage Time (Plugin / Default Tooling) & Poses Created (Plugin Only / Default Tooling Only / Both)\\
    \hline
    1 & 29 mins / 13 mins  & 0 / 0 / 3 \\
    \hline
    2 & 39 mins / 14 mins & 0 / 0 / 3 \\
    \hline
    3 & 20 mins / 17 mins & 0 / 2 / 6 \\
    \hline
    4 & 22 mins / 8 mins & 1 / 0 / 2 \\
    \hline
    5 & 28 mins / 0 mins & 3 / 0 / 0 \\
    \hline
\end{tabular}
\label{table:usageresults}
\end{table}

Table \ref{table:usageresults} tabulates the usage metrics of our plugin by subject. Times reported only account for ``useful" work done by the subject in progression towards a final pose result. Default tooling usage is largely comprised of adjusting hand joints - time spent adjusting object transforms are generally negligible. Time spent reading documentation, asking questions, drafting intermediate poses, and any other ``stall" events are omitted. Subjects averaged between 4.625 minutes and 17.667 minutes per pose, with Subject 3 being the fastest and 2 the slowest. All subjects spent more time using the plugin than existing tooling, albeit for different reasons (e.g. experimentation of capabilities, fine grained contact adjustment, creating and destroying contacts at different locations, etc.).

We examine subject 3 for an approximate tooling efficiency comparison. Subject 3 largely used the tool as a ``coarse solver" by exploiting a recurring set of contacts. The subject avoided creating new contacts per pose, and instead kept returning to a saved default set. The subject transferred contacts to the object, computed a coarse pose, and then used manual tooling to tweak the coarse estimate into the final pose. The subject required approximately an equal amount of time creating each pose using both our plugin in combination with existing FK posing tools and existing posing tools only; however, we found the plugin-augmented poses to be substantially higher quality. Therefore, we argue that our tool in its current state can be considered valuable not necessarily for raw throughput speed, but rather, as validated by qualitative assessment results, for its impact on quality.

\section{Subject-Generated Results}

\begin{figure*}
\centering
\includegraphics[width=0.95\linewidth]{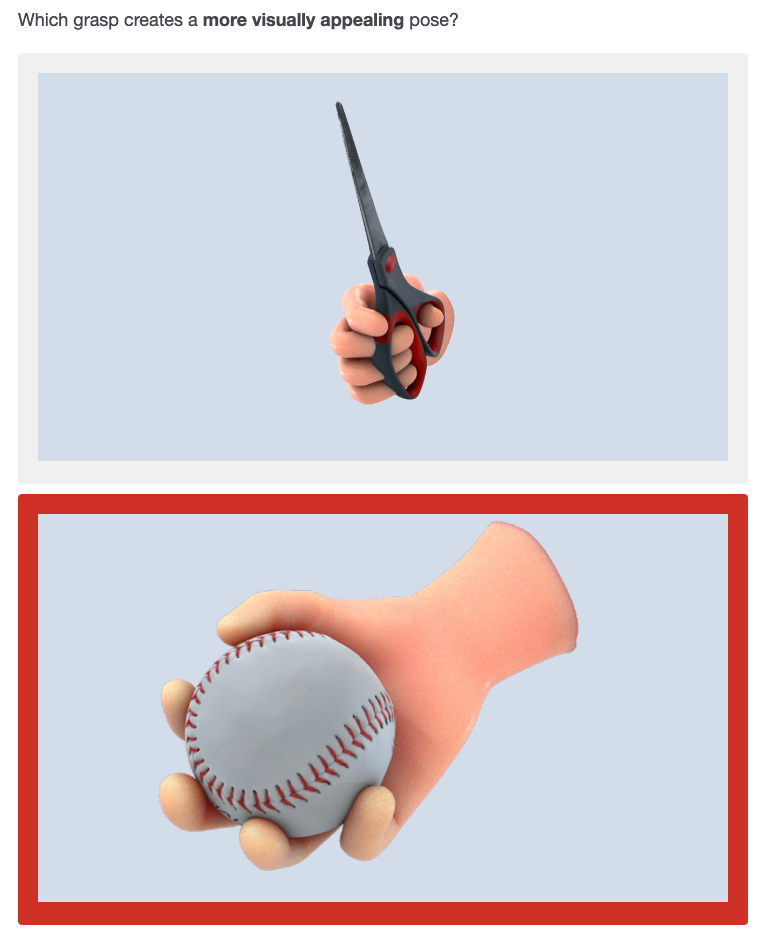}
\caption{Example qualitative assessment survey question, where the highlighted red box indicates subject selection.}
\label{fig:surveysamplequestion}
\end{figure*}

The remainder of this document contains images of poses generated by user study subjects and distributed online during qualitative assessments. Corresponding asterisk colors indicate paired A/B images in the qualitative assessment survey. Survey participants were asked to select between the two paired images in each category outlined in Section 7.2. An example question prompt is illustrated in Figure~\ref{fig:surveysamplequestion}.

\begin{figure*}
\centering
\includegraphics[width=1.0\linewidth]{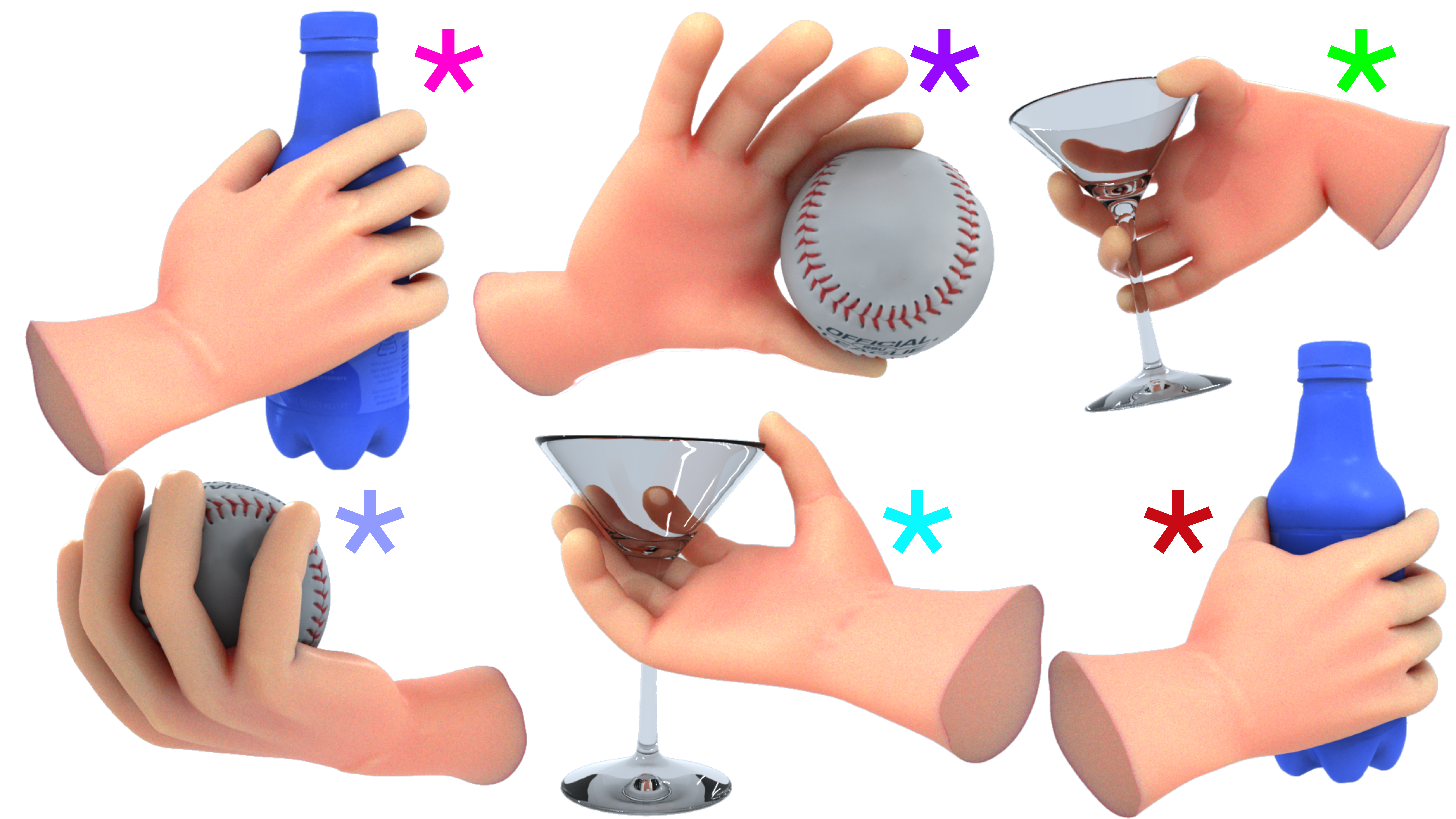}
\includegraphics[width=1.0\linewidth]{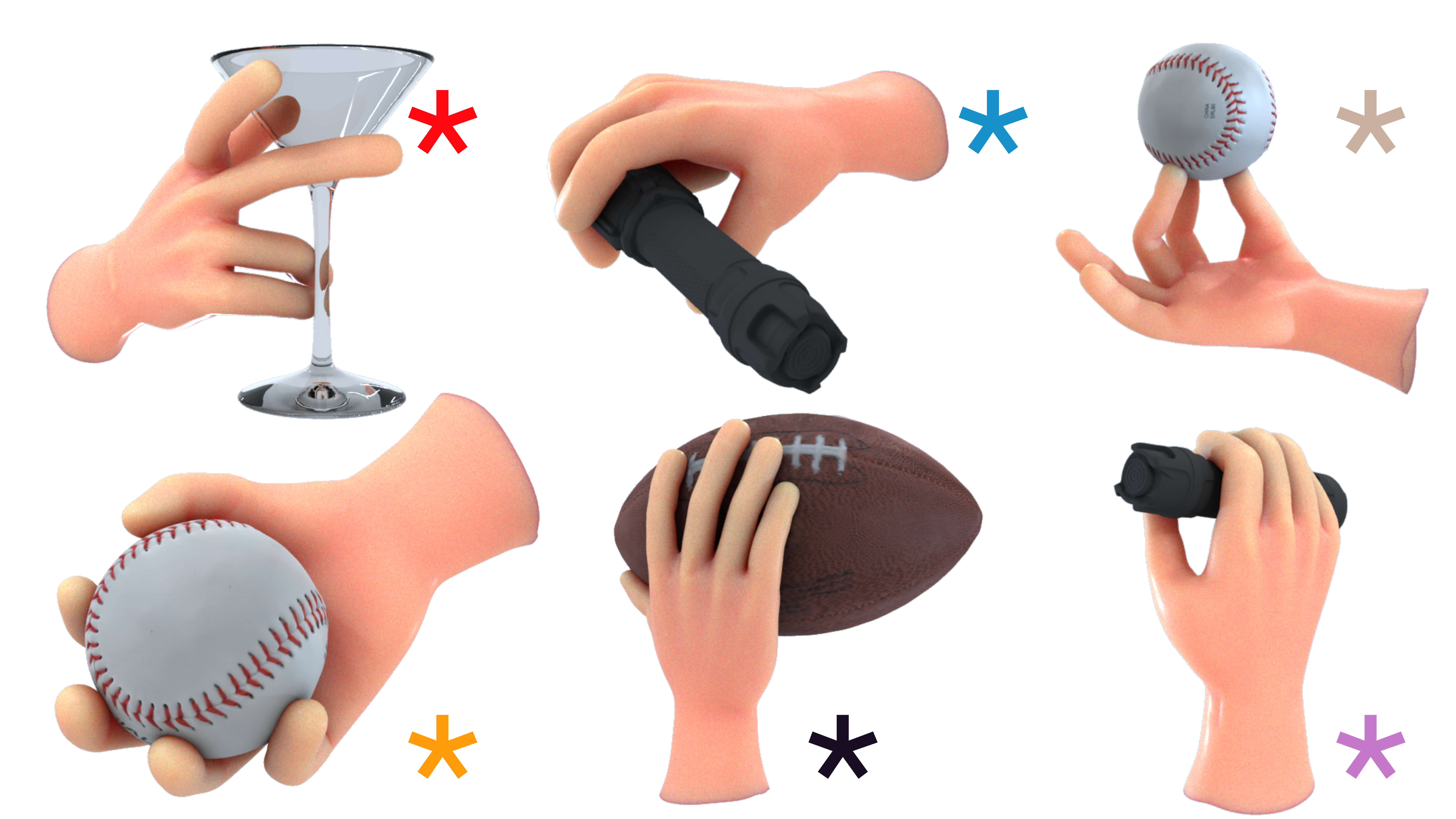}
\caption{Subject results selected for distribution in qualitative studies using our plugin in a substantial capacity.}
\label{fig:subjectcontactresults}
\end{figure*}

\begin{figure*}
\centering
\includegraphics[width=1.0\linewidth]{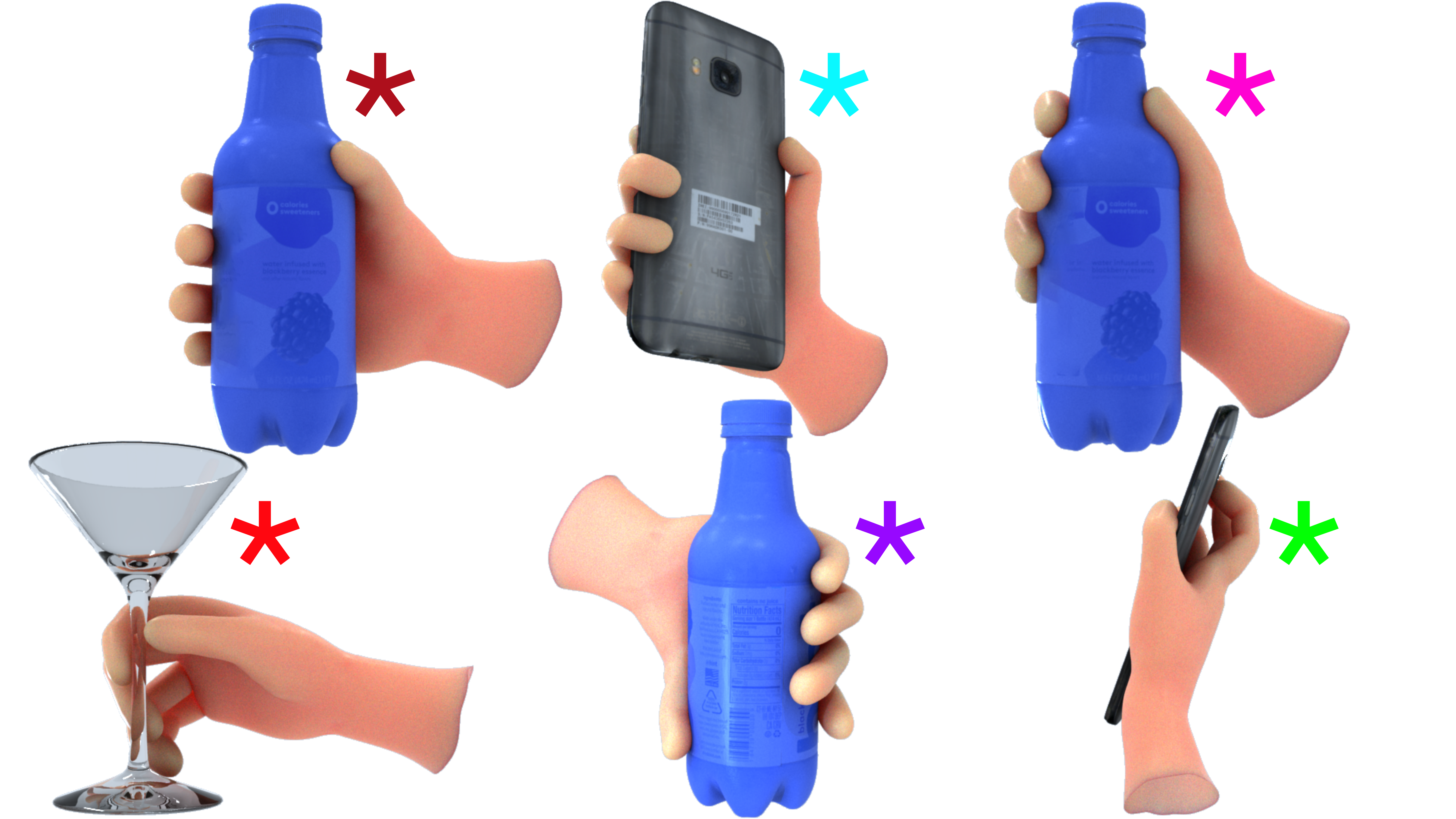}
\includegraphics[width=1.0\linewidth]{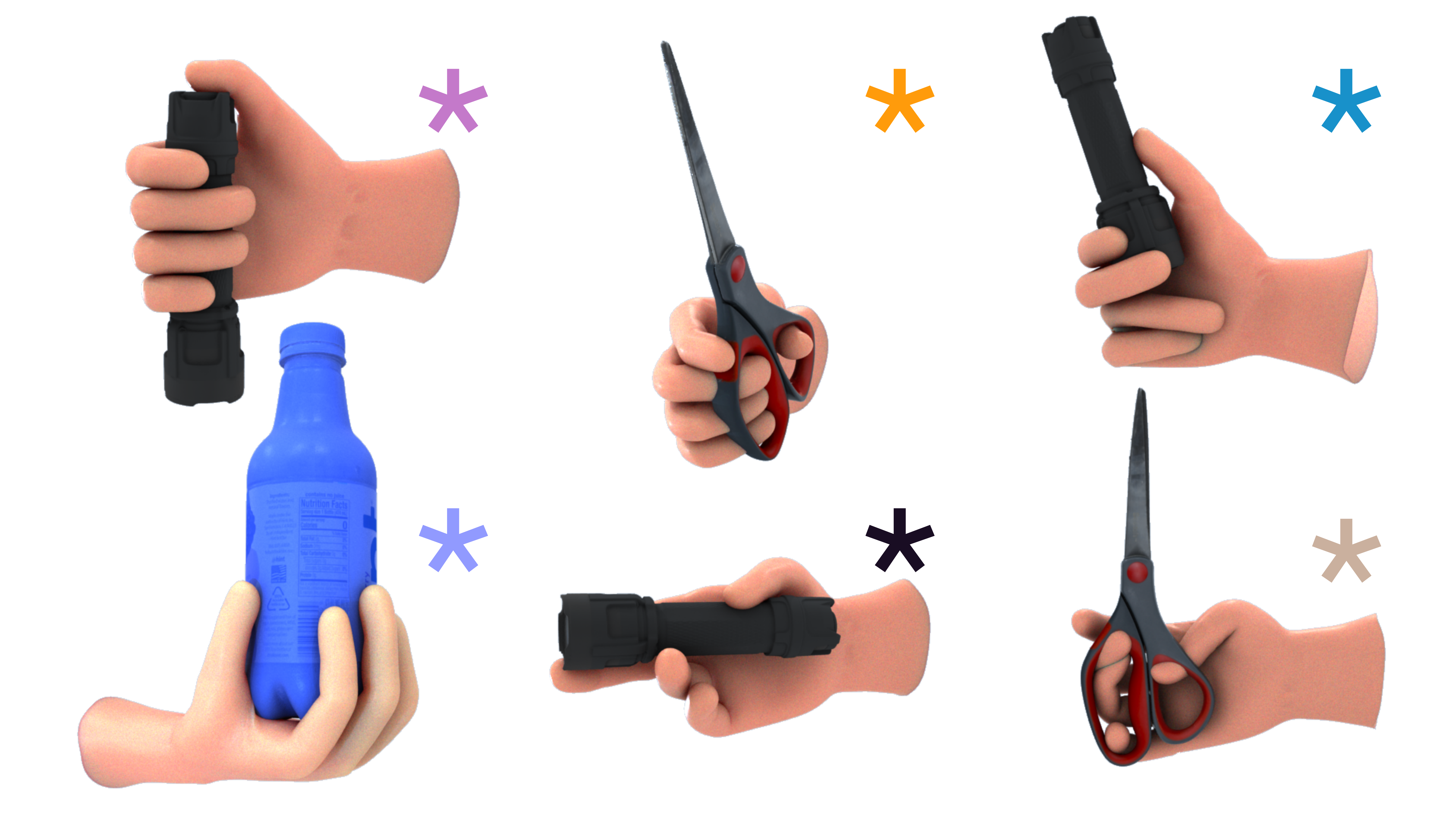}
\caption{Comparison pose results distributed in the qualitative studies survey generated using fully manual FK tooling.}
\label{fig:mayaresults}
\end{figure*}

\end{document}


\noindent We thank the reviewers for their constructive comments of our last manuscript. While we were disappointed with the outcome, we hope our revised manuscript will merit your reconsideration. We also welcome Reviewer 6, whose comments we unfortunately did not have an opportunity to address in our previous rebuttal. We refer to specific reviewers using identification numbers from the last review process for the remainder of the document.

\medskip

\noindent We have made a number of major revisions to our last manuscript. Specifically, we have added the following new items:

\begin{itemize}
    \item A new section detailing the production process overview that motivates both the design of our methods and overall project motivation
    \item Improved explanations on patch editing with additional supplementary diagrams
    \item A new section on user studies
    \item A new section on comparisons to existing contact models and posing algorithms
    \item A new section on qualitative evaluation of results
    \item A new related works subsection
    \item An appendix of pseudocode for each editing operation
    \item A secondary 9 minute supplementary video containing the tutorial distributed to subjects prior to participation in the user study, which jointly serves to illustrate both the time scales of our operations as well as provide a detailed overview of our plugin
\end{itemize}

\noindent Along with these additions, we have also revised a considerable number of figures.

\medskip

\noindent Per the suggestion of Reviewer 2, we have opted to have our new manuscript considered exclusively as a journal submission in order to move all essential information from the original supplement into the main document. We have still included a supplement with this submission, but note that it is only intended to provide additional figures.

\medskip

\noindent We hope that the aforementioned additions, as well as overall edits to the existing manuscript, sufficiently address the comments and criticisms from the last review cycle.

\medskip

\noindent The remainder of this document will target specific comments from multiple and individual reviewers:

\medskip

Regarding contact generation, please see